\documentclass[aps,prb,twocolumn,superscriptaddress,showpacs]{revtex4-1}
\usepackage{graphicx}
\usepackage{dcolumn}
\usepackage{bm}
\usepackage{color,soul}
\usepackage{amsmath}
 \usepackage{braket}
 \usepackage{csquotes}
 \usepackage{txfonts}

\begin{document}

\preprint{APS/123-QED}

\title{Role of on-site Coulomb energy and negative-charge transfer in a
Dirac semi-metal NiTe$_2$}

\author{A. R. Shelke}
\affiliation{National Synchrotron Radiation Research Center, Hsinchu 300092, Taiwan}
\author{C. -W. Chuang}
\affiliation{Experimentelle Physik VII, Universit\"{a}t W\"{u}rzburg, Am Hubland, D-97074 W\"{u}rzburg, Germany}
\author{S. Hamamoto}
\affiliation{RIKEN SPring-8 Center, Hyogo 679-5148, Japan}
\author{M. Oura}
\affiliation{RIKEN SPring-8 Center, Hyogo 679-5148, Japan}
\author{M. Yoshimura}
\affiliation{National Synchrotron Radiation Research Center, Hsinchu 300092, Taiwan}
\author{N. Hiraoka}
\affiliation{National Synchrotron Radiation Research Center, Hsinchu 300092, Taiwan}
\author{C.-N. Kuo}
\affiliation{Program on Key Materials, Academy of Innovative Semiconductor and Sustainable Manufacturing (AISSM), National Cheng Kung University, Tainan 70101, Taiwan}
\affiliation{Department of Physics, National Cheng Kung University, Tainan 70101, Taiwan}
\affiliation{Taiwan Consortium of Emergent Crystalline Materials, National Science and Technology Council, Taipei, Taiwan}
\author{C.-S. Lue}
\affiliation{Program on Key Materials, Academy of Innovative Semiconductor and Sustainable Manufacturing (AISSM), National Cheng Kung University, Tainan 70101, Taiwan}
\affiliation{Department of Physics, National Cheng Kung University, Tainan 70101, Taiwan}
\affiliation{Taiwan Consortium of Emergent Crystalline Materials, National Science and Technology Council, Taipei, Taiwan}
\author{A. Fujimori}
\affiliation{National Synchrotron Radiation Research Center, Hsinchu 300092, Taiwan}
\affiliation{Department of Physics, The University of Tokyo, 7-3-1 Hongo, Bunkyo-ku, Tokyo 113-0033, Japan}
\affiliation{Center for Quantum Technology, and Department of Physics, National Tsing Hua University, Hsinchu 300044, Taiwan}
\author{A. Chainani}
\affiliation{National Synchrotron Radiation Research Center, Hsinchu 300092, Taiwan}

\date{\today}
\begin{abstract}
Angle-resolved photoemission spectroscopy (ARPES) combined with band structure calculations have shown that the layered transition metal dichalcogenide(TMD) NiTe$_2$ is a type-II Dirac semimetal. However, conflicting conclusions were reported regarding the role of electron correlations in NiTe$_2$. We study core-levels and valence band electronic structure of single crystal NiTe$_2$ using soft and hard x-ray photoemission spectroscopy(SXPES, HAXPES), X-ray absorption spectroscopy(XAS) and Ni $2p-3d$ Resonant-PES to quantify electronic parameters in NiTe$_2$. The Ni $3d$ on-site Coulomb energy ($U_{dd}$) is quantified from measurements of the Ni $3d$ single particle density of states(DOS) and the two-hole correlation satellite. The Ni $2p$ core level and $L$-edge XAS spectra are analyzed by charge-transfer (CT) cluster model calculations using the experimental $U_{dd}$, and it shows that NiTe$_2$ exhibits a negative CT energy $\Delta$. A comparative analysis of NiO $L$-edge XAS confirms its well-known strongly correlated CT insulator character, with a larger $U_{dd}$  and positive $\Delta$. The $d$-$p$ hybridization strength $T_{eg}$ for NiTe$_2$$<$NiO, and shows that $T_{eg}$ is not responsible for reducing $U_{dd}$ in NiTe\textsubscript{2} compared to NiO. The negative-$\Delta$ and a reduced $U_{dd}$ leads to the increase in $d^n$ count on the Ni site in NiTe$_{2}$ by nearly one electron. However, importantly, since $U_{dd}$$>$$|\Delta|$, a finite repulsive $U_{dd}$ results in pushing $d$-states away from Fermi level and this is required to make NiTe$_{2}$ a moderately correlated Dirac semi-metal with band inversion in the $p$-$p$ type lowest energy excitations.
\end{abstract}

\maketitle


\section{\label{sec:levelI}Introduction}

The fascinating transport properties and electronic structure of topological insulators (TIs)  predicted by theory\cite{Murakami, Kane, Bernevig, Fu2007, Moore, Roy, Zhang} and their experimental validation\cite{Molenkamp, Hsieh2008} established the field of TIs in condensed matter. In particular, (i) transport measurements of a Quantum Spin Hall insulator proved existence of a pair of gapless helical edge-states in HgTe/(Hg,Cd)Te quantum wells\cite{Molenkamp}, and (ii) ARPES showed Dirac-type metallic surface state bands inside a bulk charge gap induced by spin-orbit coupling (SOC) with band inversion in Bi$_{1-x}$Sb$_x$\cite{Hsieh2008}. While inverted band gaps in substituted semiconductors leading to Dirac/Weyl type bands in the gap was recognized earlier\cite{Volkov,Pankratov}, Dirac cones and band inversions were soon reported in several materials like Bi$_2$Se$_3$, Sb$_2$Te$_3$, etc. \cite{Xia2009, Chen2009, Hsieh2009} confirming their TI character.

Similarly, the 3-dimensional (3D) Weyl semimetal (WSM) was predicted to show a band structure with a pair of Dirac/Weyl nodes of opposite chirality, separated in momentum($k$)-space\cite{BB}. WSMs were also predicted to show a Drude weight vanishing as T$^2$ and it was confirmed by experiments\cite{Sushkov}. ARPES studies showed characteristic Fermi arcs connecting the Dirac/Weyl nodes in TaAs\cite{Lv,Huang} and NbP\cite{Souma}. The WSMs require broken time-reversal symmetry and/or broken inversion symmetry\cite{Zyuzin}. In contrast, the Dirac semimetals(DSMs) form in the presence of time-reversal and inversion symmetries\cite{Burkov,Bahramy}. They are of 2 types: Type I DSMs obey Lorentz symmetry and show regular Dirac-type linear dispersions along all 3 directions in $k$-space as  seen in Na$_3$Bi\cite{Liu, Xu} and Cd$_3$As$_2$\cite{Neupane, Borisenko}; Type II DSMs violate Lorentz symmetry and show tilted Dirac cones with a Dirac point at which an electron and a hole pocket touch each other\cite{Burkov,Bahramy}. 

Many layered TMDs have  shown Type II DSM behavior\cite{Bahramy, Zhang2017, Noh, Yan, Clark, Chakraborty2023, Xu2018, Ghosh2019, Mukherjee, hlevyack2021, Nurmamat, Fischer, Bhatt_2025} and in this work, we focus on the electronic structure of NiTe$_2$. ARPES studies showed a bulk Type-II Dirac point in NiTe$_2$ lying $\sim$20-68 meV above $E_F$ by three groups\cite{Ghosh2019, Mukherjee, hlevyack2021}, while another group reported it to lie 150 meV below $E_F$\cite{Nurmamat}. NiTe$_2$ also shows surface states at $E_F$, and another Dirac-like crossing lies far ($\sim$1.5 eV) below $E_F$\cite{Ghosh2019, Mukherjee, hlevyack2021, Nurmamat, Bhatt_2025}.
However, while three studies on NiTe$_2$ using ARPES and band structure calculations had ruled out correlation effects in NiTe$_2$\cite{Ghosh2019, Mukherjee, hlevyack2021}, two very recent studies have concluded the importance of correlations for the Te $5p$ derived bands in NiTe$_2$\cite{Fischer, Bhatt_2025}. Fischer et al.\cite{Fischer} carried out DFT calculations and calculated many-body effects within the GW approximation. They found improvement in the Te $5p$ bands with an increase in Dirac carrier velocity exceeding 100\% and emphasized the subtle influence of electronic interactions on band structure. Bhatt et al.\cite{Bhatt_2025} used LDA+U (with U = 5 eV) and concluded that the topological Te $5p$ surface states lying $\sim$1.5 eV below $E_F$ can be described only by including U. Surprisingly, there is no discussion of the role of correlation effects in Ni $3d$ states of NiTe$_2$ to date. 

Since Ni is divalent in NiTe$_2$ as the Te atoms are dimerized (Te$_2^{2-}$)\cite{Jobic, bensch1996}, it is important to consider the example of divalent NiO which is known to be a strongly correlated CT insulator. Thus, Ni$^{2+}$ states in NiTe$_2$ may also be correlated, and we use the case of NiO as a benchmark to compare it with correlation effects in NiTe$_2$. However, it is known that NiTe$_2$ is a Pauli paramagnetic metal\cite{Zheng} and may as well be less correlated than NiO. On the other hand, bulk sensitive dHvA experiments showed that 
individual bands had to be shifted in a somewhat ad-hoc manner (one with an energy shift of -60 meV, and another with a shift of +100 meV) to match with measured dispersions. Also, calculated and measured values of light and nearly isotropic effective masses showed discrepancies. It was concluded that DFT calculations failed to capture the finer details of the  band structure of NiTe$_2$\cite{Zheng}.

While materials like NiTe$_2$ are called semimetals because they show small electron- and hole-pockets, their actual conductivities are quite high due to high carrier mobilities. For example, NiTe$_2$ exhibits a very high conductivity of $\sigma$(T = 2 K)$\sim$1.0$\times$10$^6$S/m\cite{Shi}. Magnetic susceptibility of NiTe$_2$ does not show evidence for magnetic ordering down to T = 2 K, but is Pauli paramagnetic as mentioned above\cite{Mao, Zheng}. It is also known that the NiTe$_{2-x}$ series exhibits pressure-induced superconductivity\cite{8}. A giant Josephson diode effect in Josephson junctions formed using NiTe$_2$ was also reported recently\cite{Pal}. NiTe$_{2-x}$ nanomaterials also work as electrocatalysts for hydrogen evolution reaction\cite{Shi,10}. In a very recent study, it was shown that NiTe$_{2}$ is a topological superconductor with a very low $T_C$ = 261 mK\cite{CHe}. Thus, NiTe$_2$ is important in terms of fundamental and applied properties. 

In this work, we carry out SXPES, HAXPES, XAS and Ni $2p-3d$ resonant PES to determine electronic parameters in NiTe$_2$. We quantify $U_{dd}$ in the Ni $3d$ states using experimental Ni $3d$ single particle DOS and the two-hole correlation satellite. A comparative analysis using CT cluster model calculations for $L$-edge XAS shows that $U_{dd}$ = 3.7 eV and $\Delta$ = -2.8 eV for NiTe$_2$,  while  $U_{dd}$ = 7.0 eV and $\Delta$ = 6.0 eV for NiO. The obtained values of $d$-$p$ hybridization strength $T_{eg}$ indicate it is smaller for NiTe$_2$ compared to NiO. Since $U_{dd}$$>$$|\Delta|$, it indicates the important role of a moderate repulsive $U_{dd}$ in making NiTe$_{2}$ a topological metal with \textit{p}$\rightarrow$\textit{p} type lowest energy excitations in the Zaanen-Sawatzky-Allen (ZSA) scheme \cite{zaanen1985}.

\section{\label{sec:levelI}Methods}

\textbf{Synthesis and structure characterization:} Single crystals of NiTe$_2$ were prepared by the chemical vapor transport method, using iodine as the transport agent\cite{Prodan}. High-purity Ni (99.95\%) and Te (99.999\%) powders were mixed with iodine, sealed in an evacuated quartz tube, and heated for 10 days in a two-zone furnace. Finally, the quartz tube was quenched into an ice-water bath from the growth temperature of 800$^\circ$C. The obtained single crystals are hexagonal in shape with typical dimensions of 4 mm$\times$4 mm$\times$0.5 mm. The crystal structure was characterized using powder X-ray diffraction (XRD) (Bruker D2 phaser diffractometer) with Cu-K$\alpha$ radiation. The single crystal quality was confirmed and crystallization directions were identified by the Laue diffraction method (Photonic Science). 
The XRD results confirmed the 1T-CdI$_2$-type trigonal structure (space group of P$\bar{3}$m1 (No. 164)) with the flat-surfaces corresponding to the (001) plane. The obtained lattice parameters of NiTe$_2$ are $a = b$ = 3.853$\AA$ and $c$ = 5.260$\AA$. These values are very close to reported values of $a = b$ = 3.843$\AA$ and $c$ = 5.266$\AA$\cite{TMdatabase}.

\textbf{Electron spectroscopy:}
HAXPES core levels and valence band measurements were carried out at BL-12XU, Taiwan Beamline in SPring-8, Japan, using linearly polarized x-ray beam with incident photon energy $h\nu$ = 6.5 keV. Liquid He closed cycle cryostat was used to cool down the sample to T = 25 K. The $E_F$ of Au thin film was measured at 25 K to calibrate the binding energy (BE) scale and determine the total energy resolution ( $\Delta$E = 270 meV) from a fit to the Au film Fermi edge. NiTe$_2$ single crystal was cleaved using a top-post in ultra-high vacuum (UHV) at 5.5 $\times$ 10$^{-9}$ mbar in the preparation chamber and then quickly transferred to the main chamber at 7.0 $\times$ 10$^{-10}$ mbar for the measurements. Soft X-ray PES (SXPES) core level and valence band, Ni $L_{3,2}$-edge XAS and 2$p-3d$ resonance PES measurements were carried out at BL-17SU, RIKEN beamline in SPring-8, Japan, using a circularly polarized X-ray beam. Ni $L_{3,2}$-edge XAS spectra of single crystal NiO was measured as a reference compound to confirm the photon energy scale calibration and for comparison with NiTe$_2$. The XAS measurements were carried out in total electron yield mode. 
SXPES core levels and valence band measurements were carried out with incident photon energy $h\nu$ = 1.5 keV. A Liquid N$_2$ flow-type cryostat was used to cool the sample down to 80 K. The total energy resolution for SXPES was set to $\Delta$E = 220-400 meV for h$\nu$ = 700 eV - 1.5 keV, as obtained from fits to the Au Fermi edge ($E_F$) measured with $h\nu$ = 700 eV and 1.5 keV at T = 80 K. Sample was cleaved in the main chamber at a UHV better than 1.0 $\times$ 10$^{-10}$ mbar.

\textbf{Cluster-model calculations:}
The Ni $L$$_{3,2}$-edge XAS and Ni 2$p$ PES core level spectra were calculated using a charge transfer multiplet cluster model with the QUANTY code\cite{haverkort2012, lu2014, haverkort2014}. A Ni$L_6$ cluster with divalent  Ni$^{2+}$ ion (3$d^8$) and 6 ligand($L$) atoms of Te in octahedral symmetry ($O_h$) was used to calculate the spectra. The initial state consists of a linear combination of $\ket{3\textit{d}^{\textit{8}}}$, $\ket{3\textit{d}\textsuperscript{9}\underline{\textit{L}}\textsuperscript{1}}$ and $\ket{3\textit{d}\textsuperscript{10}\underline{\textit{L}}\textsuperscript{2}}$ states, where \underline{\textit{L}} corresponds to hole in ligand states. 
The ground-state is the lowest energy state obtained by diagonalizing the Hamiltonian matrix in terms of the basis states $\ket{3\textit{d}^{\textit{8}}}$, $\ket{3\textit{d}\textsuperscript{9}\underline{\textit{L}}\textsuperscript{1}}$ and $\ket{3\textit{d}\textsuperscript{10}\underline{\textit{L}}\textsuperscript{2}}$ states. The results also provide the coefficients of the basis states in the ground state. The contributions of each basis state in the ground state is then obtained as the square of its coefficient. 
A check on the validity of the calculations is confirmed by the sum of the percentages which add up to 100\%.
The value of $U_{dd}$ was fixed to 3.7 eV as obtained from the Cini-Sawatzky analysis, while $\Delta$, the hybridization strength ($T_{eg}$ and $T_{2g}$ = $T_{eg}$/2) and the crystal field splitting 10$Dq$ were varied to obtain calculated spectra close to experimental spectra. The calculated spectra were obtained by convoluting the discrete final states by broadening it with a Lorentzian function for lifetimes and a Gaussian function for the experimental spectral width. 

\section{\label{sec:levelII}results and discussion} 

\begin{figure}
\includegraphics[scale=0.3]{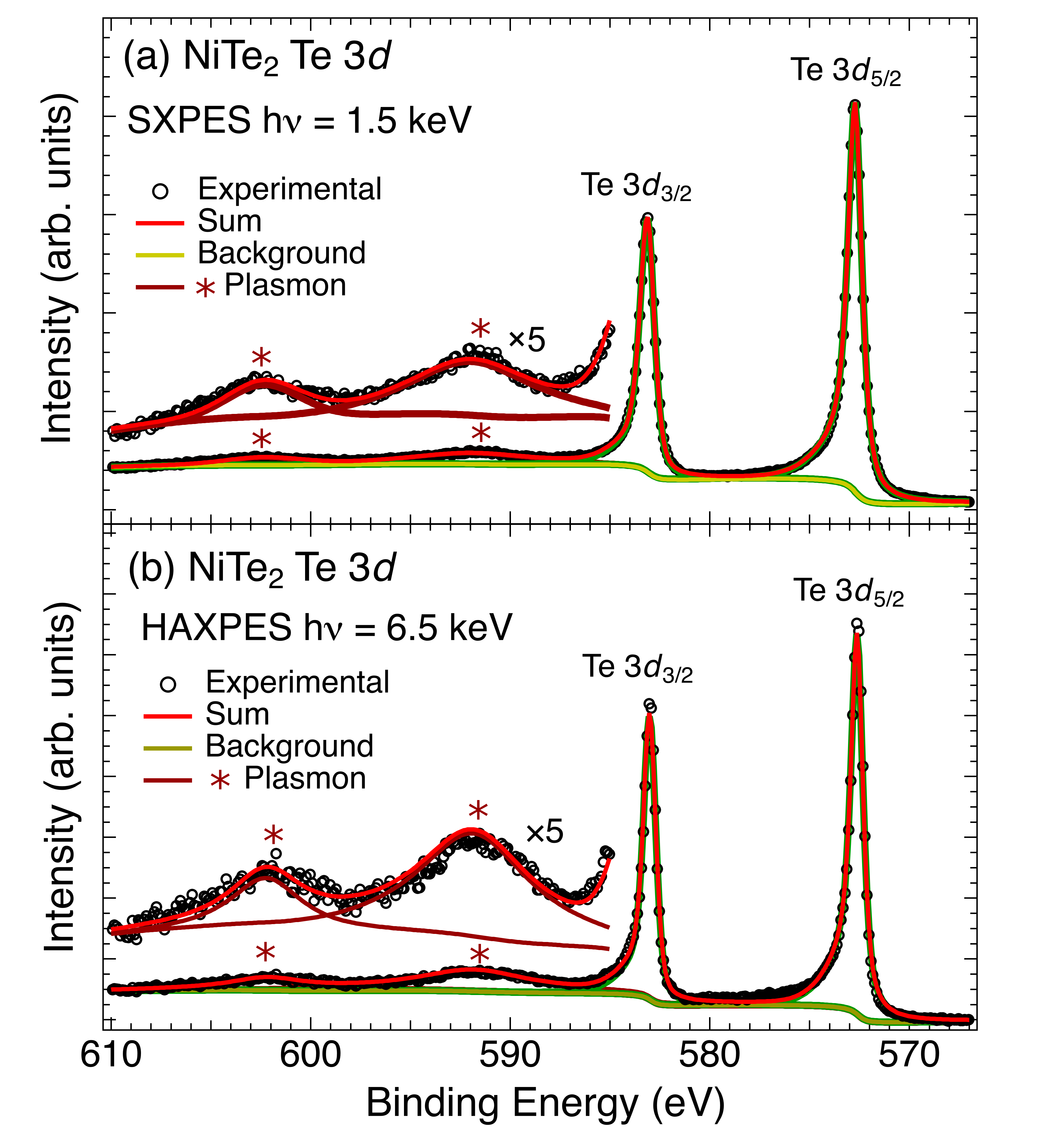}
\caption{Least-squares fitting of Te 3$d$ core levels of NiTe$_2$ single-crystal measured at $T$ = 80 K with $h\nu$ = 1.5 keV (SXPES) and at $T$ = 20 K with $h\nu$ = 6.5 keV (HAXPES)}
\end{figure}

\begin{table}[t!]
	\begin{center}
	\caption{Fitting parameters of Te 3$d$ core levels of NiTe$_2$ single-crystal measured using SXPES and HAXPES techniques}
\begin{ruledtabular}
\begin{tabular}{ccc}
 Fit component  & Binding Energy   & FWHM \\
 SXPES &          (eV) &                   (eV) \\
\hline
Te 3$d_{5/2}$ & 572.72 & 0.82 \\
Te 3$d_{3/2}$ & 583.13 & 0.88 \\
Te 3$d_{5/2}$ Plasmon & 591.90 & 7.5\\
Te 3$d_{3/2}$ Plasmon & 602.32 & 4.5\\
\hline
HAXPES \\
Te 3$d_{5/2}$ & 572.62 & 0.72 \\
Te 3$d_{3/2}$ & 583.01 & 0.74 \\
Te 3$d_{5/2}$ Plasmon & 591.82 & 7.0\\
Te 3$d_{3/2}$ Plasmon & 602.21 & 4.5\\
\end{tabular}
\end{ruledtabular}
\end{center}
\end{table}

\par Figures 1(a) and (b) show the Te 3$d$ core levels of NiTe$_2$ measured using SXPES and HAXPES, respectively.
The spectra show two sharp single peaks for Te 3$d_{5/2}$ and Te 3$d_{3/2}$ main peaks and weak satellite features at higher BEs. A least-squares fitting to the Te 3$d_{5/2}$ and Te 3$d_{3/2}$ main peaks and weak features is superimposed as full lines on the experimental spectra (empty circles). The main peaks were fitted with single asymmetric Voigt functions, and the weak features could be fitted with symmetric Gaussian functions. The peak BEs and FWHMs are listed in Table I. The Te 3$d_{5/2}$ and Te 3$d_{3/2}$ main peaks are at 572.72 eV$\pm$ 0.1 eV and at 583.13 eV$\pm$ 0.1 eV in SXPES and  HAXPES spectra. The observed BE values are very consistent with reported values of Te 3$d$ core levels of NiTe$_2$ \cite{bensch1996, nappini2020}. The sharp single main peaks of Te 3$d_{5/2}$ and Te 3$d_{3/2}$ and absence of any feature $\approx$3.0 eV above the main peaks indicates absence of oxidation in SXPES and HAXPES spectra\cite{nappini2020}. Further, the broad low intensity features are both positioned at $\sim$19.7 eV$\pm$0.1 eV higher BEs than the main peaks and suggests they originate from plasmon excitations. We confirm their plasmonic origin by measuring Te 3$p$ SXPES and HAXPES core levels which also show weak satellite features at $\sim$19.7 eV$\pm$0.1 eV higher BEs from the main peaks in Fig. 2, as discussed in the following.


\begin{figure}
\includegraphics[scale=0.32]{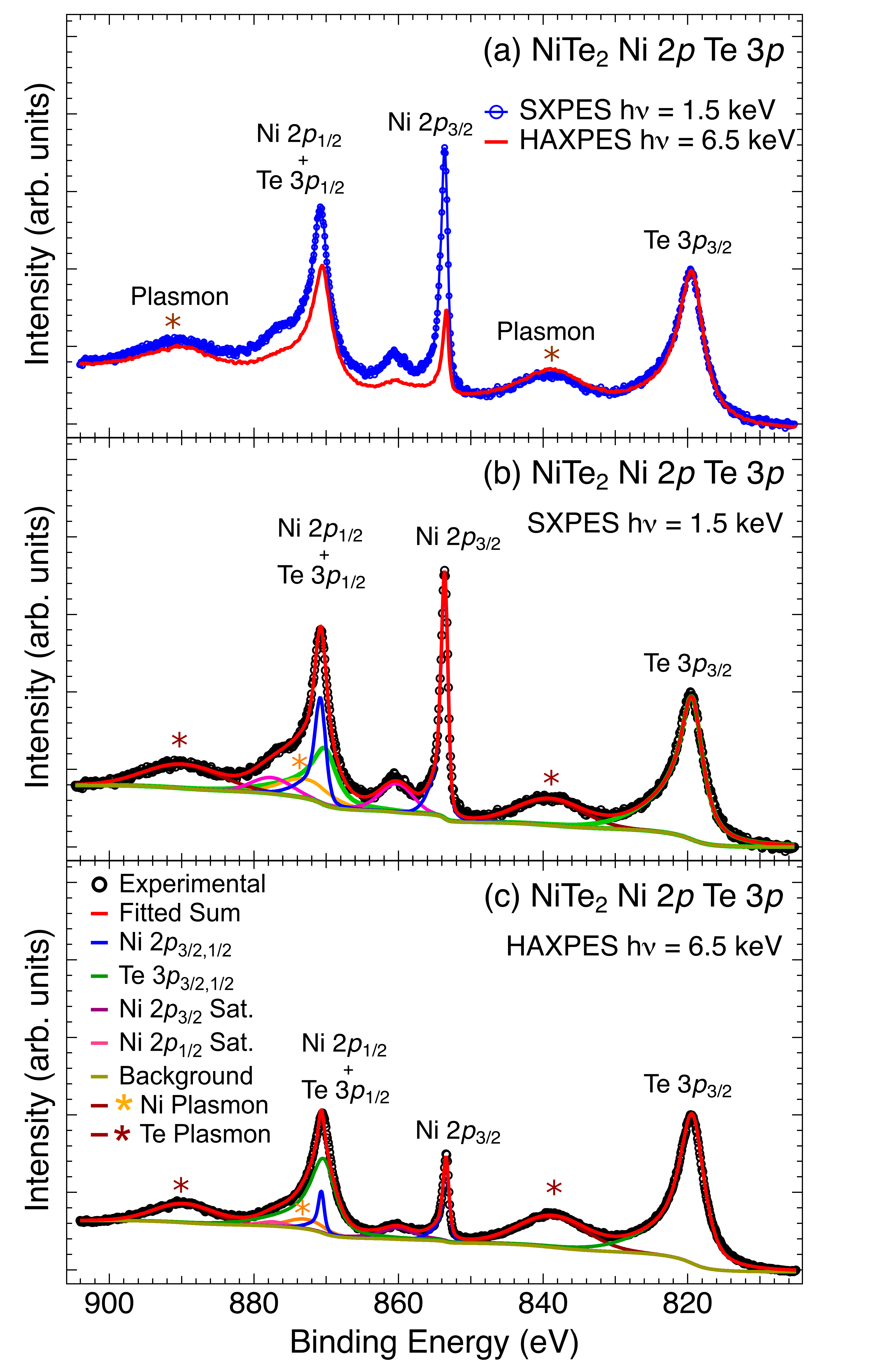}
\caption{(a) Ni 2\textit{p} and Te 3$p$ core level spectra of NiTe$_2$ single-crystal measured at $T$ = 80 K with $h\nu$ = 1.5 keV (SXPES) and at $T$ = 20 K with $h\nu$ = 6.5 keV (HAXPES). The spectra are normalized at the Te 3${p_3/2}$ main peak. (b) Least squares fitting of Ni 2$p$ and Te 3$p$ core levels of NiTe$_2$ measured with  $h\nu$ = 1.5 keV (SXPES)  and (c) Least squares fitting of Ni 2$p$ and Te 3$p$ core levels of NiTe$_2$ measured with $h\nu$ = 6.5 keV (HAXPES).}
\end{figure}

\begin{table}[t!]
	\begin{center}
	\caption{Fitting parameters of Ni 2$p$ and Te 3$p$ core levels of NiTe$_2$ single-crystal measured using SXPES and HAXPES techniques}
\begin{ruledtabular}
\begin{tabular}{ccc}
 Fit component  & Binding Energy   & FWHM \\
 SXPES &          (eV) &                   (eV) \\
\hline
Ni 2$p_{3/2}$ & 853.62 & 1.10  \\
Ni 2$p_{1/2}$ & 870.86 & 1.70  \\
Te 3$p_{3/2}$ & 819.44 & 3.5 \\
Te 3$p_{1/2}$ & 870.24 & 3.5 \\
Ni 2$p_{3/2}$ Satellite &860.30 & 5.18\\
Ni 2$p_{1/2}$ Satellite &877.51 & 6.5\\
Ni 2$p_{3/2}$ Plasmon & 872.80 & 8.1\\
Te 3$p_{3/2}$ Plasmon & 838.66 & 10.82\\
Te 3$p_{1/2}$+Ni 2$p_{1/2}$Plasmon & 889.78 & 12.39\\
\hline
HAXPES \\
Ni 2$p_{3/2}$ & 853.43 & 0.97  \\
Ni 2$p_{1/2}$ & 870.63 & 0.97  \\
Te 3$p_{3/2}$ & 819.40 & 3.93 \\
Te 3$p_{1/2}$ & 870.20 & 3.93 \\
Ni 2$p_{3/2}$ Satellite &860.10 & 5.5\\
Ni 2$p_{1/2}$ Satellite &877.32 & 4.5\\
Ni 2$p_{3/2}$ Plasmon & 872.78 & 6.0\\
Te 3$p_{3/2}$ Plasmon & 838.66 & 11.19\\
Te 3$p_{1/2}$+Ni 2$p_{1/2}$ Plasmon & 889.72 & 9.76\\
\end{tabular}
\end{ruledtabular}
\end{center}
\end{table}

Figure 2(a) shows the Ni 2$p$ and Te 3$p$ core level range from 812 - 900 eV BE  measured with HAXPES ($h\nu$ = 6.5 keV) and SXPES ($h\nu$ = 1.5 keV). The spectra show three high intensity peaks instead of the expected four peaks:  Ni 2$p_{3/2}$, Ni 2$p_{1/2}$, Te 3$p_{3/2}$ and Te 3$p_{1/2}$, together with weak features. The spectra are normalized at the lowest BE peak at 819.4 eV BE which corresponds to the Te 3$p_{3/2}$ main peak. The next higher BE main peak has a BE of 853.6 eV and corresponds to the Ni 2$p_{3/2}$ peak. These values of   Te 3$p_{1/2}$ and  Ni 2$p_{3/2}$ are consistent with known BEs of metallic Te and Ni\cite{XPShandbook}. Based on this information and the known spin-orbit (SO) doublet splittings of Ni 2$p$ ($\sim$17.3 eV) and Te 3$p$ ($\sim$51.0 eV) from reference data\cite{XPShandbook}, we can expect the SO split Ni 2$p_{1/2}$ peak at a BE of 870.9 eV, and the Te 3$p_{1/2}$ peak at 870.4 eV. Thus the Ni 2$p_{1/2}$ peak and the Te 3$p_{1/2}$ peak are expected to be separated just by 0.5 eV, and hence, they essentially overlap each other. Further, as the HAXPES and SX-PES spectra have been normalized at the Te 3$p_{3/2}$, it can be seen in Fig. 2(a) that the  Ni 2$p_{3/2}$ as well as the overlapping Ni 2$p_{3/2}$+Te 3$p_{1/2}$ peaks show higher intensities compared to Te 3$p_{3/2}$ peak in SX-PES compared to HAXPES. This is due to the fact that Ni 2$p$ states have a higher photo-ionization cross-section (PICS) than Te 3$p$ states for the photon energies used in SXPES compared to HAXPES\cite{trzhaskovskaya2018}. This observation suggests that the Ni 2$p_{1/2}$ peak and the Te 3$p_{1/2}$ peak overlap and we confirm it by carrying out a least-squares fitting using discussed below.

In order to characterize the details of the independent contributions of Ni 2$p_{1/2}$ and Te 3$p_{1/2}$ main peaks, as well as that of the weaker features, 
we carried out a least-squares fitting of the spectra, as shown in Fig. 2(b) and (c) for SXPES and HAXPES data, respectively. The fitting used SO splitting energies of 17.3 eV and 50.8 eV for Ni 2$p$ and Te 3$p$, respectively, consistent with reported values for metallic Ni and Te\cite{XPShandbook}. The Te 3$p_{3/2}$ and Te 3$p_{1/2}$ shows broad satellites at $\approx$19.7 $\pm$ 0.1 eV, and similar features at $\approx$19.7 higher BEs are also seen in Te $3d$ spectra (Fig. 1), we assign them to plasmon excitations. For the fitting, we constrained the intensity ratios of Ni 2$p_{3/2}$ : Ni 2$p_{1/2}$ and Te 3$p_{3/2}$ : Te 3$p_{1/2}$ to the expected value of 2 : 1 and used asymmetric Doniach-Sunjic Voigt lineshapes for the main peaks and charge-transfer satellites, and symmetric Gaussians for the plasmon features.  However, since the Ni 2$p_{3/2}$ main peak shows a satellite about 8 eV higher BE, it indicates the Ni 2$p$ peaks also show a charge transfer(CT) satellite similar to other known Nickel compounds. Accordingly, just above the Ni 2$p_{3/2}$, we used an asymmetric Voigt function for the CT satellite while just above the Ni 2$p_{1/2}$ main peak, we included one Gaussian for the Ni 2$p_{3/2}$ main peak plasmon  and another asymmetric Voigt function for the weaker CT satellite.
The total and component fits to the Ni 2$p$ and Te 3$p$ spectra for both SXPES and HAXPES are shown in Fig. 2(b) and (c), respectively, and the obtained BEs and their full-width and half-maximum (FWHM) are listed in Table II. The main peak BEs are very similar for Ni 2$p$ and Te 3$p$ core levels in both, SXPES and HAXPES data. In particular, the Ni 2$p_{1/2}$ and Te 3$p_{1/2}$ peak positions are quite close to each other (0.5 eV$\pm$0.1 eV) as estimated from the BEs using Ni 2$p_{3/2}$ and Te 3$p_{3/2}$ BEs and the known SO splitting.
The Ni 2$p$ and Te 3$p$ results are also consistent with a recent measurement\cite{nappini2020}) on NiTe$_2$. Further, the present Te 3$p$ results indicate the main peak BEs and the plasmon energies are quite close to values reported by us for CoTe$_2$ measurements\cite{ShelkeCoTe2}.

\begin{figure*}
    \centering
    \includegraphics[width=1\textwidth]{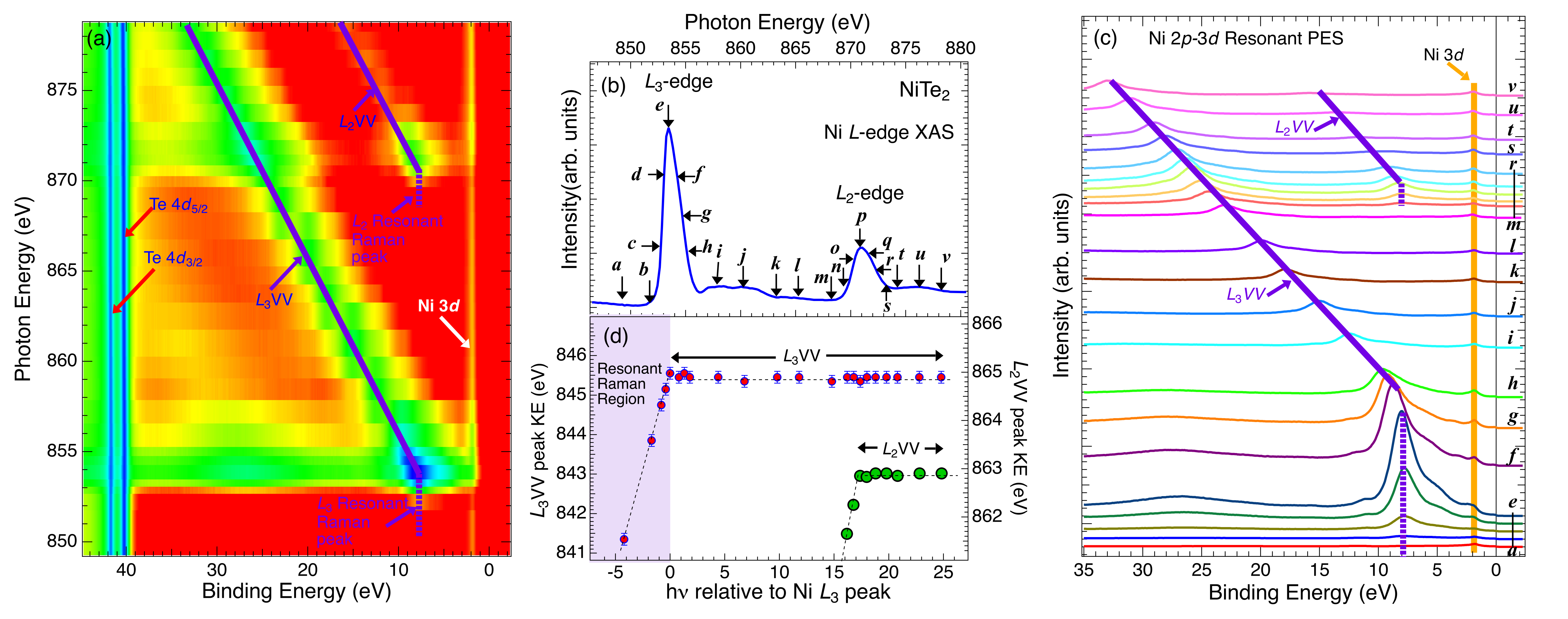}
    \caption{(a). The Ni 2$p$-3$d$ resonant-PES valence band intensity map plotted as a function of incident photon energies ($h\nu$ = 849-879 eV) versus BE  ( = -1.2 to 45.8 eV). (b) The Ni $L_3$- and $L_2$-edge XAS plotted as a function of $h\nu$ (top X-axis). (c) Valence band spectra (BE = -1.2 to 35.0 eV) measured at select $h\nu$ values (labelled $a - v$) across the $L_3$- and $L_2$-edges of Fig. 1(b). (d) The kinetic energy of the Resonant Raman peak which becomes the $L_3VV$ Auger peak, plotted as a function of $h\nu$ relative to the XAS $L_3$ peak energy(bottom X-axis).}
    \label{fig:img1}
\end{figure*}

We then carried out Ni 2$p$ - 3$d$ resonant-PES of NiTe$_2$ to quantify the experimental value of $U_{dd}$ from measurements of the single particle 3$d$ partial density of states (PDOS) and the two hole correlation satellite using the Cini-Sawatzky method \cite{cini1976, cini1977, sawatzky1977}. Fig. 3(a) shows the Ni $2p - 3d$ resonant PES valence band intensity map as a function of photon energies ($h\nu$ = 849-879 eV) versus BE ( = -1.2 to 45.8 eV). The resonant PES map is obtained by first measuring the Ni $L_{3,2}$-edge XAS Fig. 3(b) as a function of $h\nu$ (top X-axis). Then, at different incident photon energies $a -v$ marked by arrows in the XAS spectrum(Fig. 3(b)), we measure the valence band spectra and plot the spectral intensity as the map shown in Fig. 3(a). Fig. 3(c) shows valence band spectra (BE = -1.2 eV to 35.0 eV) measured at select $h\nu$ values (labelled $a - v$) across the $L_3$ and $L_2$ edge of Fig. 3(b). The BEs are calibrated with respect to $E_F$ of metallic NiTe$_2$ and the spectra are normalized to the shallow Te 4$d_{5/2, 3/2}$ core level peaks (bright blue vertical lines at $\sim$40 and 42 eV BEs) in the map. The normalization shown in Fig. 3a and 3c at the Te 4$d_{5/2,3/2}$ peaks is carried out to emphasize resonance effects in the main Ni $3d$ states near 
$h\nu$$=$853 eV, as well as evolution of the two-hole correlation satellite peak for $h\nu$ $>$853 eV, discussed below.
Several well-resolved spectral features derived from Ni $d$ bands and Te 5$p$ bands are observed in the valence band spectra. The weak spectral feature at 1.90 eV BE shows a small increase in intensity when incident $h\nu$ values just cross the $L_3$ and $L_2$ edges and indicate a Ni 2$p - 3d$ resonance of the Ni 3$d$ derived partial density of states (PDOS) ( green and yellow vertical line observed in Fig. 3(a) map, and yellow line also marked in Fig. 3(c)), consistent with low energy He I ARPES measurements of Ni 3d band\cite{orders1982}. The central feature in the Fig. 3(a) map is the diagonal from right bottom to left top. It  originates in a feature at 7.90 eV BE in the map and Fig. 3(c), and shows a significantly large increase in intensity on increasing  $h\nu$ from $a - e$, i.e., upto the Ni $L_3$-edge maxima (blue vertical dotted line in the map and Fig. 3(c). On increasing $h\nu$ further from $f - l$, this feature systematically moves to higher BEs with a shift equal to the increase in $h\nu$. This behavior of shift in BE equal to increase in $h\nu$ indicates its Ni $L_3VV$ Auger feature (marked by the blue diagonal line in the map and in Fig. 3(c)). Fig. 3(d) shows a plot of the kinetic energy of this peak as a function of $h\nu$ relative to the XAS $L_3$ peak energy (bottom X-axis). The corresponding incident $h\nu$ values are the same as top X-axis of Fig. 3(b). This indicates that the $\sim$7.90 eV constant BE feature for $h\nu$ = $a - e$ represents a resonant Raman feature and it becomes a correlation satellite with two holes in the final state\cite{guillot1977, weinelt1997, hufner2000}. The map and Fig. 3(c) also shows a weak broad feature at $\approx$19.3 eV BE for $h\nu$ across the Ni $L_3$ edge, which is attributed to a plasmon feature as seen in core level spectra. For higher $h\nu$ from $m - v$, the map and Fig. 3(c) again show a weak resonant Raman behavior(blue vertical dotted line), followed by the $L_2VV$ two-hole correlation satellite Auger peak (blue full line) of the 7.90 eV BE feature. The corresponding kinetic energy of this peak as a function of $h\nu$ is also plotted in Fig. 3(d) and confirms its  Auger character. 

\begin{figure}
\includegraphics[scale=0.34]{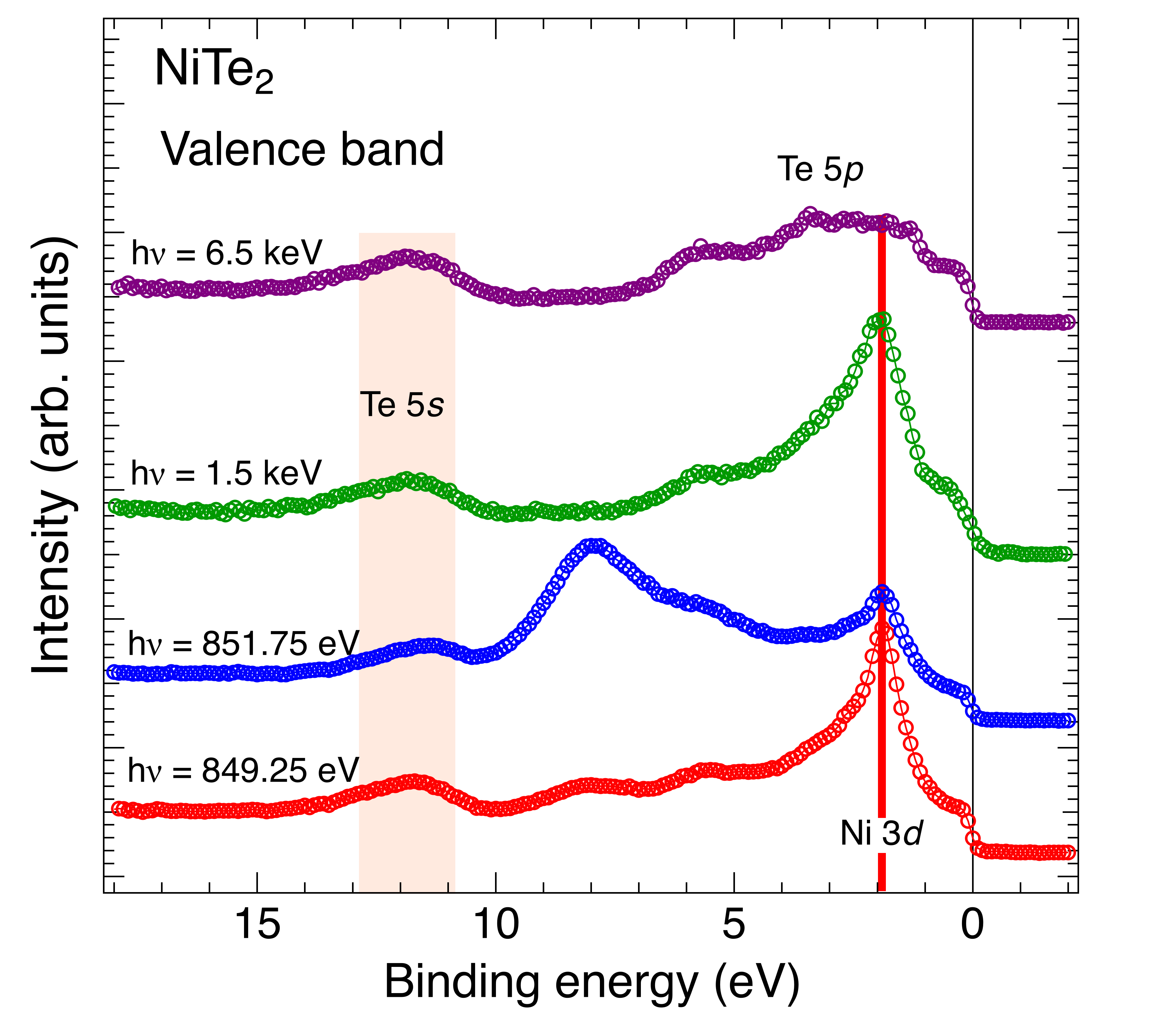}
\caption{NiTe$_2$ valence band measurements at different off-resonant $h\nu$ values of $h\nu$ = 849.25 eV, 1.5 keV, 6.5 keV, and one in the Resonant Raman region with $h\nu$ = 851.75 eV.}
\end{figure}

Figure 4 shows the valence band spectra measured at three different off-resonant photon energies of values $h\nu$ = 849.25 eV,1.5 keV, 6.5 keV, and one in the Resonant Raman region with $h\nu$ = 851.75 eV. All these spectra are normalized at the broad spectral feature at 11.5 eV BE, which corresponds to the Te 5$s$ shallow core level. The purpose of this normalization is to emphasize the relative spectral weights of Te 5$p$ states and Ni 3$d$ states. The normalization in Fig. 4 shows the intensities of the Ni 3$d$ main band as well as the two-hole satellite at 7.9 eV BE relative to the intensity of the Te 5$s$ states. It also identifies the Te 5$p$ states in the $h\nu$ = 6.5 keV HAXPES spectrum. Please note that for photon energy $b$, $h\nu$ = 851.75 eV spectrum (blue curve), the photon energy is in the Resonant Raman region and the resonance effect is strong for $\sim$8 eV BE, but weak for the main Ni 3$d$ band at 1.9 eV BE. . This happens because photon energy $b$, h$\nu$ = 851.75 eV, is still below the main Ni $L_3$ peak energy (see Fig. 3b, 3d).

Comparison of the spectra obtained with the two lowest  $h\nu$ values, indicates that the Ni 3$d$ PDOS peak at 1.90 eV BE in Fig. 4 shows a weak suppression in the Resonant Raman region ($h\nu$($b$) = 851.75 eV) but the two-hole satellite at 7.9 eV BE gets enhanced, when compared to the off-resonant ($h\nu$($a$) = 849.25 eV) spectrum. However, the spectral shape recovers at higher energies from photon energy $h\nu$($e$) = 853 eV (Figs. 3a, 3c). The spectral shape for off-resonant $h\nu$ =1.5 keV is also fairly similar to off-resonant $h\nu$ = 849.25 eV. On the other hand, at $h\nu$ = 6.5 keV, the 1.90 eV BE and 7.9 eV BE features get strongly suppressed at $h\nu$ = 6.5 keV due to the strongly reduced PICS of Ni 3$d$ states compared to the Te 5$p$ states at $h\nu$ = 6.5 keV\cite{trzhaskovskaya2018}.  At $h\nu$ = 6.5 keV, the Te 5$p$ states dominate the spectrum as the PICS ratio of Te 5$p$ : Ni 3$d$ is 6.41 \cite{trzhaskovskaya2018}. 
This indicates that the feature at E$_F$ corresponds to dominantly Te $p$ PDOS. This is consistent with reported DFT electronic structure calculations and ARPES measurements which concluded that Te $p$ states lie at $E_F$ and within 0.5 eV BE \cite{Ghosh2019, Mukherjee, hlevyack2021, Nurmamat, settembri2024, Bhatt_2025}. As we will show in Fig. 6 below, this observation is also consistent with a negative $\Delta$ in NiTe$_2$. Before that, we quantify $U_{dd}$ using the Cini-Sawatzky method\cite{cini1976, cini1977, sawatzky1977} applied to the experimental Ni 3$d$ PDOS and two-hole correlation satellite data. 

\begin{figure}
\includegraphics[scale=0.34]{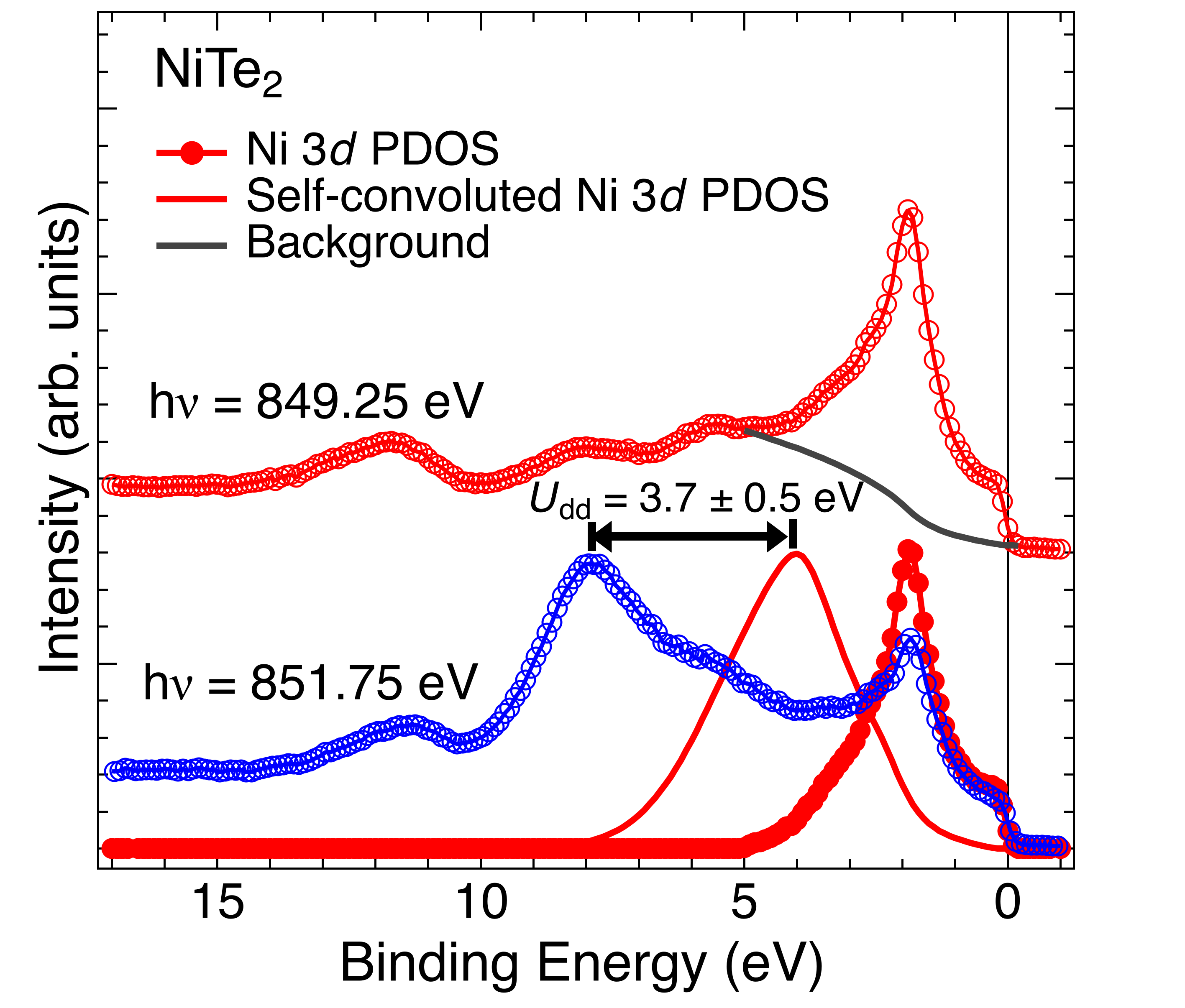}
\caption{The Ni 3$d$ partial density of states (red color;$\medbullet$) are obtained by subtracting a Shirley-type integral background (black line) from off-resonance valence band spectrum obtained at $h\nu$ =  849.25 eV (red color;$\medcirc$). Comparing the peak in the self-convolution of Ni 3$d$ PDOS (red line) with the two-hole correlation satellite peak seen in the Resonant Raman region spectrum ($h\nu$ = 851.75 eV; blue color $\medcirc$), the average on-site Coulomb energy is estimated as the energy difference between the two peaks.}
\end{figure}

Figure 5 shows the off resonance spectrum obtained with $h\nu$ = 849.25 eV before/after (red $\medcirc$/$\medbullet$) subtracting a Shirley-type integral background in order to separate out the single-particle Ni 3$d$ PDOS from the Te 5$p$ states at higher BEs. 
We have also tested for a linear background and the results showed hardly any change for the self-convolved two-hole spectrum as well as the subsequent determination of $U_{dd}$. 
The single-particle PDOS was then numerically self-convoluted to obtain the two-hole spectrum (red line), and its peak energy represents the average two-hole energy without correlations. This two hole spectrum was then compared with the spectrum exhibiting the experimental two-hole correlation satellite spectrum in the resonant Raman region ($h\nu$ = 851.75 eV; blue color $\medcirc$). In the Cini-Sawatzky method, the energy difference between the peak in the two-hole spectrum without correlations and the peak of the experimental two-hole correlation satellite gives an estimate of the on-site Coulomb energy $U_{dd}$. We obtain the value of $U_{dd}$ = 3.7 eV for Ni 3$d$ states, indicating that the NiTe$_2$ is a moderately correlated metal. In the following, we use the obtained $U_{dd}$ value in charge-transfer cluster model calculations to simulate the Ni 2$p$ core-level PES and $L$-edge XAS spectra of NiTe$_2$ to independently check its validity.

\begin{figure}
\includegraphics[scale=0.34]{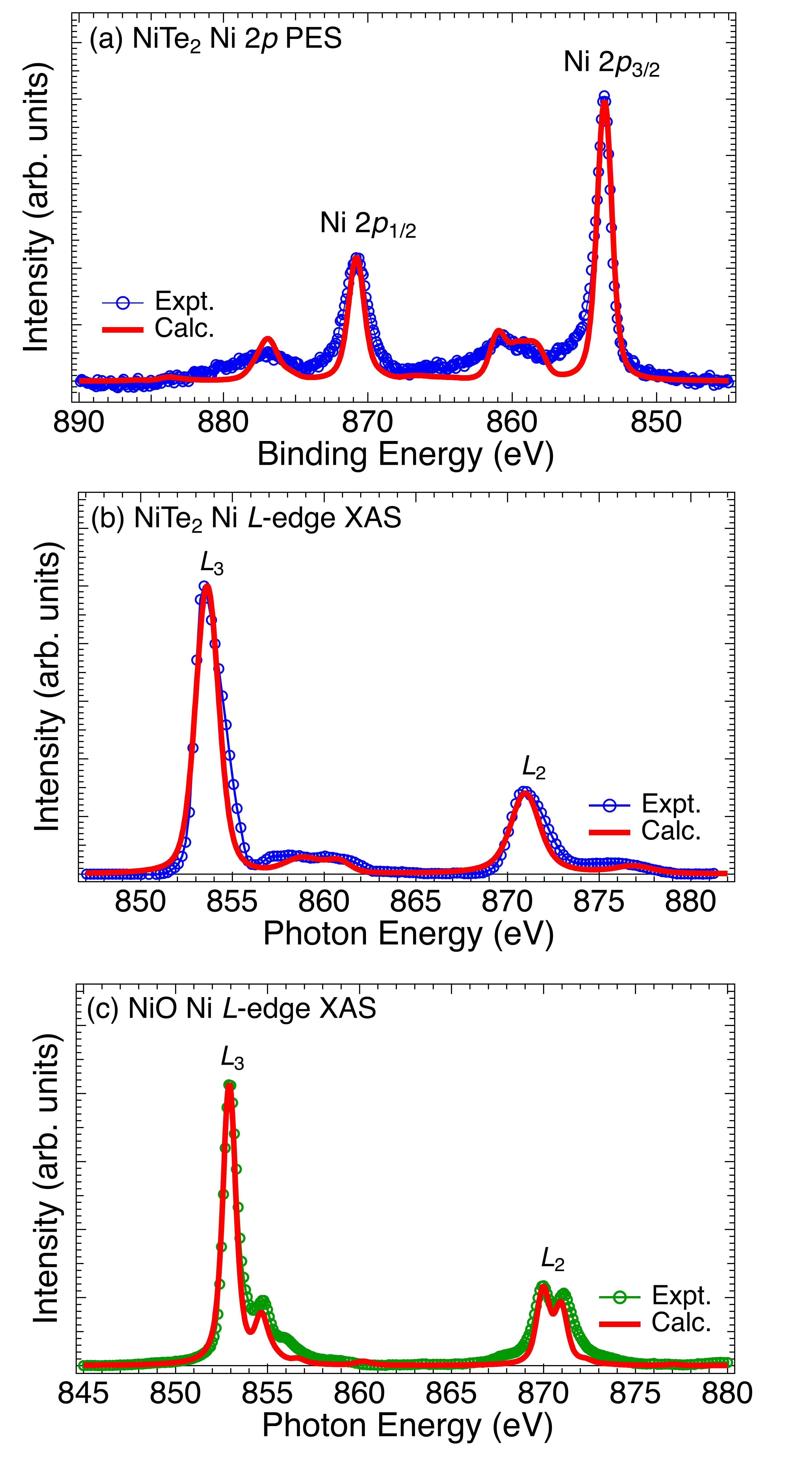}
\caption{(a) Ni 2\textit{p} SXPES core level, (b) Ni $L_{3,2}$-edge XAS of NiTe$_2$, and (c) Ni $L_{3,2}$-edge XAS of NiO, compared with atomic multiplet charge transfer cluster model calculations.}
\end{figure}

Figures 6(a) and (b) show the experimental Ni 2$p$ PES core level and Ni $L$-edge XAS spectra of NiTe$_2$ compared with charge transfer cluster model simulations obtained using $U_{dd}$ = 3.7 eV and optimizing other electronic parameters. Due to the overlapping Ni 2$p_{1/2}$ and Te 3$p_{1/2}$ peaks, the experimental Ni 2$p$ PES core level spectrum shown in Fig. 6(a) was obtained by subtracting out the fits to the Te 3$p_{1/2}$ main peak and its plasmon, as well as the Ni 2$p_{3/2}$ plasmon fit.
For the Ni $L_{3,2}$-edge XAS shown in Fig. 6(b), since the Te M$_{5}$-edge region did not show significant intensity
in the XAS spectrum due to very low cross-section, we did not need to subtract a corresponding feature for the Te M$_{4}$-edge region. 
We confirmed that the spectral lineshape of Ni $L_{3,2}$-edge XAS shown in Fig. 6(b) is very similar to the reported Ni $L_{3,2}$-edge XAS of NiTe$_2$ single-crystal cleaved surface, including the satellite structures for the Ni $L_{3,2}$-edge \cite{nappini2020}. We then carried out an extensive set of calculations by varying $\Delta$, $T_{eg}$, $T_{t2g}$ (=$T_{eg}$/2) and 10 $Dq$ to optimize the parameters, as shown in Figs. 7 and 8 and discussed below.  Using the experimental $U_{dd}$ = 3.7 eV, and the same set of optimized parameters as listed in Table III, we could obtain calculated spectra close to the experiment for both Ni 2$p$ PES and Ni $L$-edge XAS spectra as shown in Fig. 6(a) and (b), respectively.
Similarly, using $U_{dd}$ = 7.0 eV (from earlier XAS calculations for NiO; ref.\cite{Veenendaal, alders1998}), we optimized the other parameters for Ni $L$-edge XAS spectrum of NiO (as shown in Fig. 9 and described below). In Table III, we list the optimal values used to obtain the calculated spectra close to experiment shown in Fig. 6(a,b) for NiTe$_2$ and in Fig. 6(c) for NiO.

\begin{table}[t!]
	
	 \caption{Electronic parameters, spin magnetic moments $m_S$, and total $d^n$-counts for NiTe$_2$ and NiO using  3-basis state cluster model calculations.}
\begin{tabular}{ccc}

\hline
~~~~Parameter~~~~& ~~~~NiTe$_2$~~~~&~~~~NiO~~~~\\
\hline
$U_{dd}$(eV)		&~3.7~&~7.0~ \\
$\Delta$(eV)		&~-2.8~&~6.0~\\

$T_{eg}$(eV)		&~1.8~&~2.4~\\
$T_{t2g}$(eV)		&~0.9~&~1.2~\\

$10Dq$(eV)		&~0.5~&~1.65~\\

$F_k$,$G_k$		&~1.0~&~0.8~\\

$m_S$($\mu_B$)		&~0.96~&~1.82~\\

$d^n$ count		&~9.1~&~8.14~\\

\hline
\end{tabular}
	
\end{table}

The comparison of electronic parameters for NiTe$_2$ and NiO, spin magnetic moments, and $d^n$ counts in Table III indicates that NiTe$_2$ is a moderately correlated negative charge-transfer metal in the Zaanen-Sawatzky-Allen (ZSA) scheme \cite{zaanen1985}, while NiO is a positive charge-transfer insulator as is well-known\cite{FujimoriNiO, VeenendaalPRL, Veenendaal, alders1998, Ghiasi}.
Further, we confirmed that the ground state of NiO consists of a mixture of 85.0\%  $d^8$, 14.6\% $d^{9}\underline{L}^1$ and a very small contribution of 0.4\% $d^{10}\underline{L}^2$ weights. This is consistent with the early study of Fujimori and Minami which showed NiO has a ground state consisting of a mixture of 83.0\%  $d^8$ and 17.0\% $d^{9}\underline{L}^1$ weights\cite{FujimoriNiO}. Thus NiO shows a dominantly $d^n$ character ground state.
In contrast, the ground state of NiTe$_2$ consists of a mixture of 19.5\%  $d^8$, 57.4\% $d^{9}\underline{L}^1$ and a 23.1\% $d^{10}\underline{L}^2$ weights.
The negative-$\Delta$ for NiTe$_2$ also implies that $E_F$ lies in the ligand states and the lowest energy excitations are  \textit{p}$\rightarrow$\textit{p} type excitations, consistent with ARPES results\cite{Ghosh2019, Mukherjee, hlevyack2021, Nurmamat, settembri2024, Bhatt_2025}.

\begin{figure}
\includegraphics[scale=0.20]{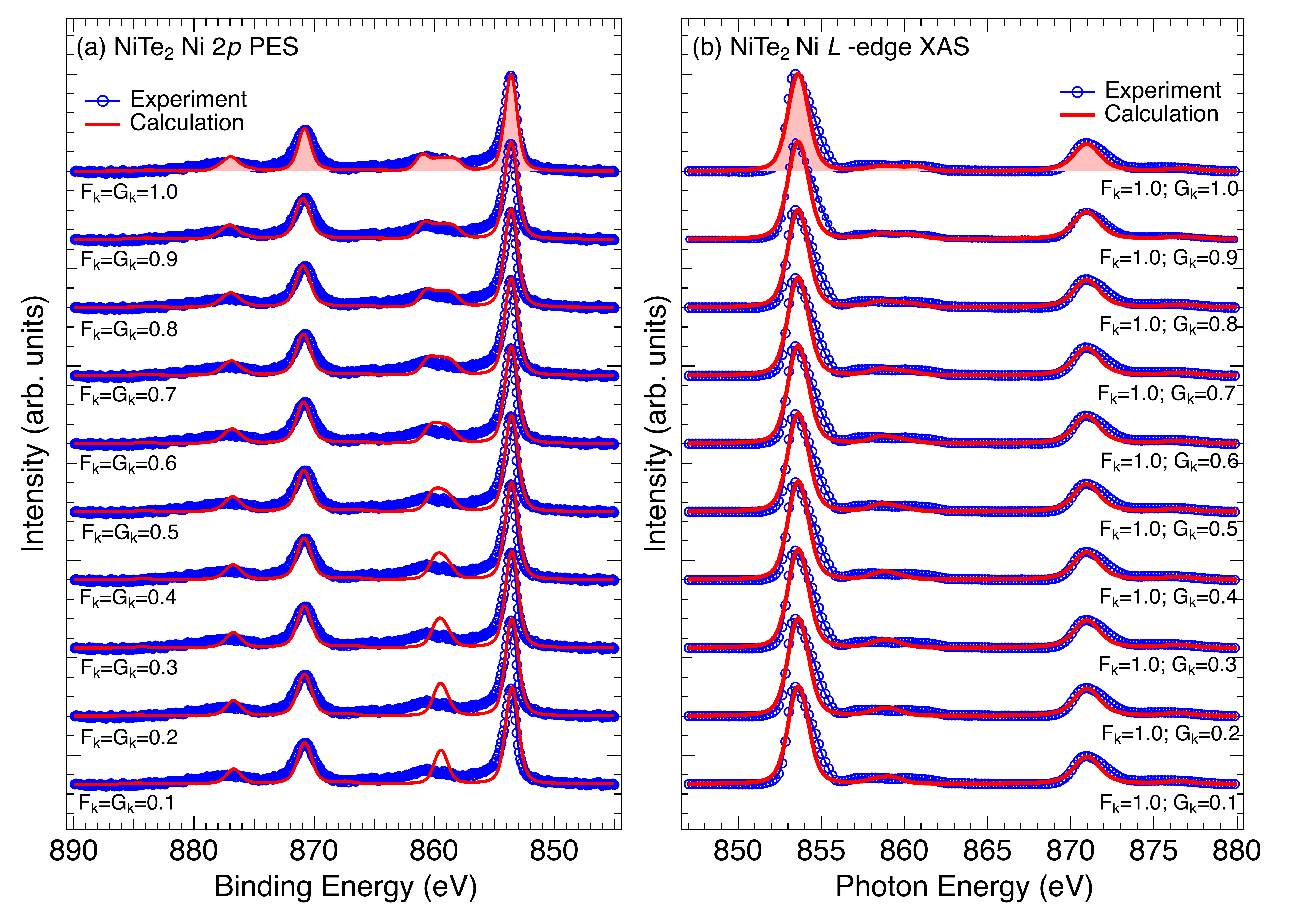}
\caption{Optimization of F\textsubscript{k} and G\textsubscript{k} for (a) Ni 2$p$ PES and (b) Ni $L$-edge XAS of NiTe$_2$}
\end{figure}
\begin{figure}
\includegraphics[scale=0.20]{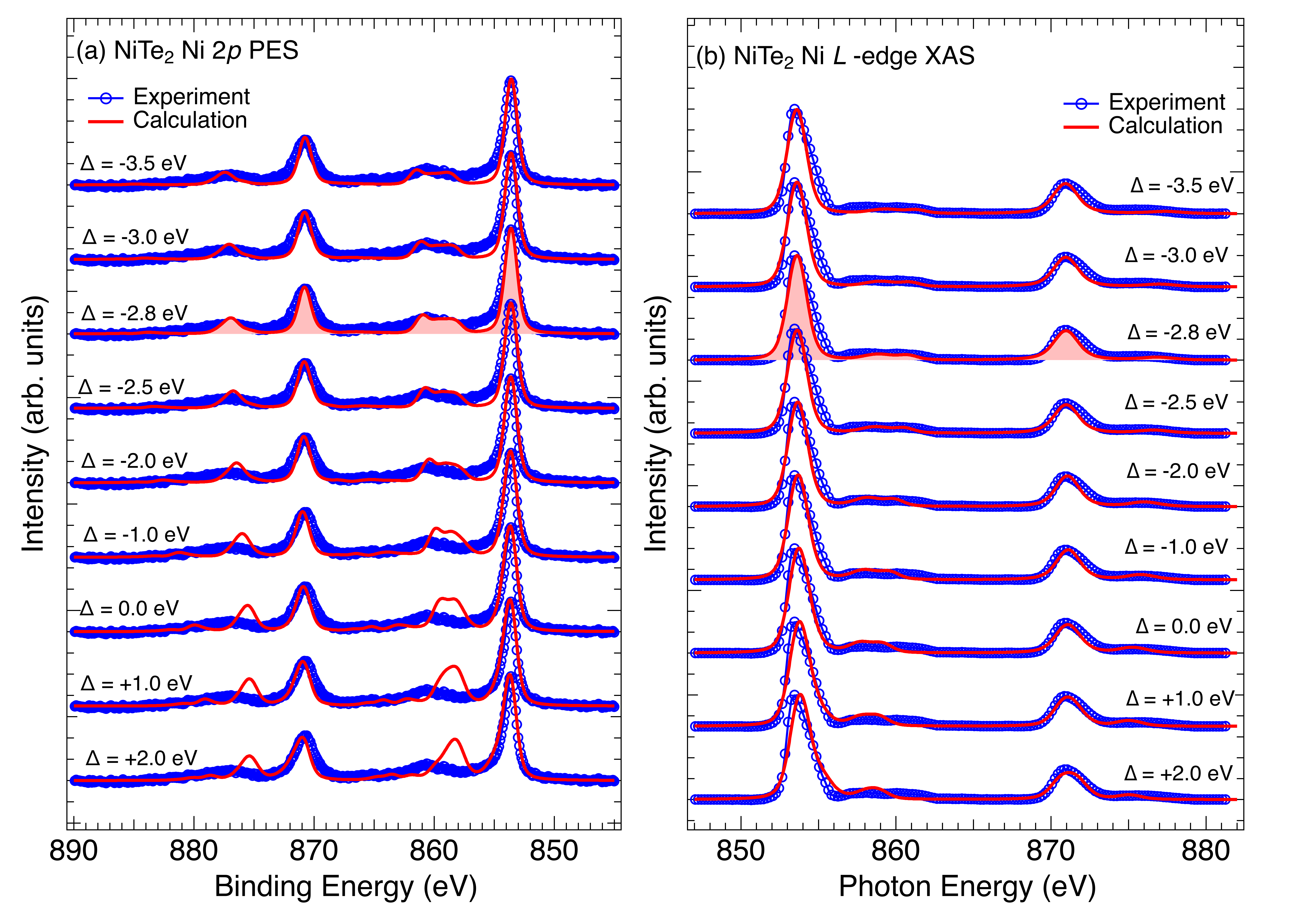}
\caption{Optimization of charge transfer energy $\Delta$ for (a) Ni 2$p$ PES and (b) Ni $L$-edge XAS of NiTe$_2$}
\end{figure}
\begin{figure}
\includegraphics[scale=0.20]{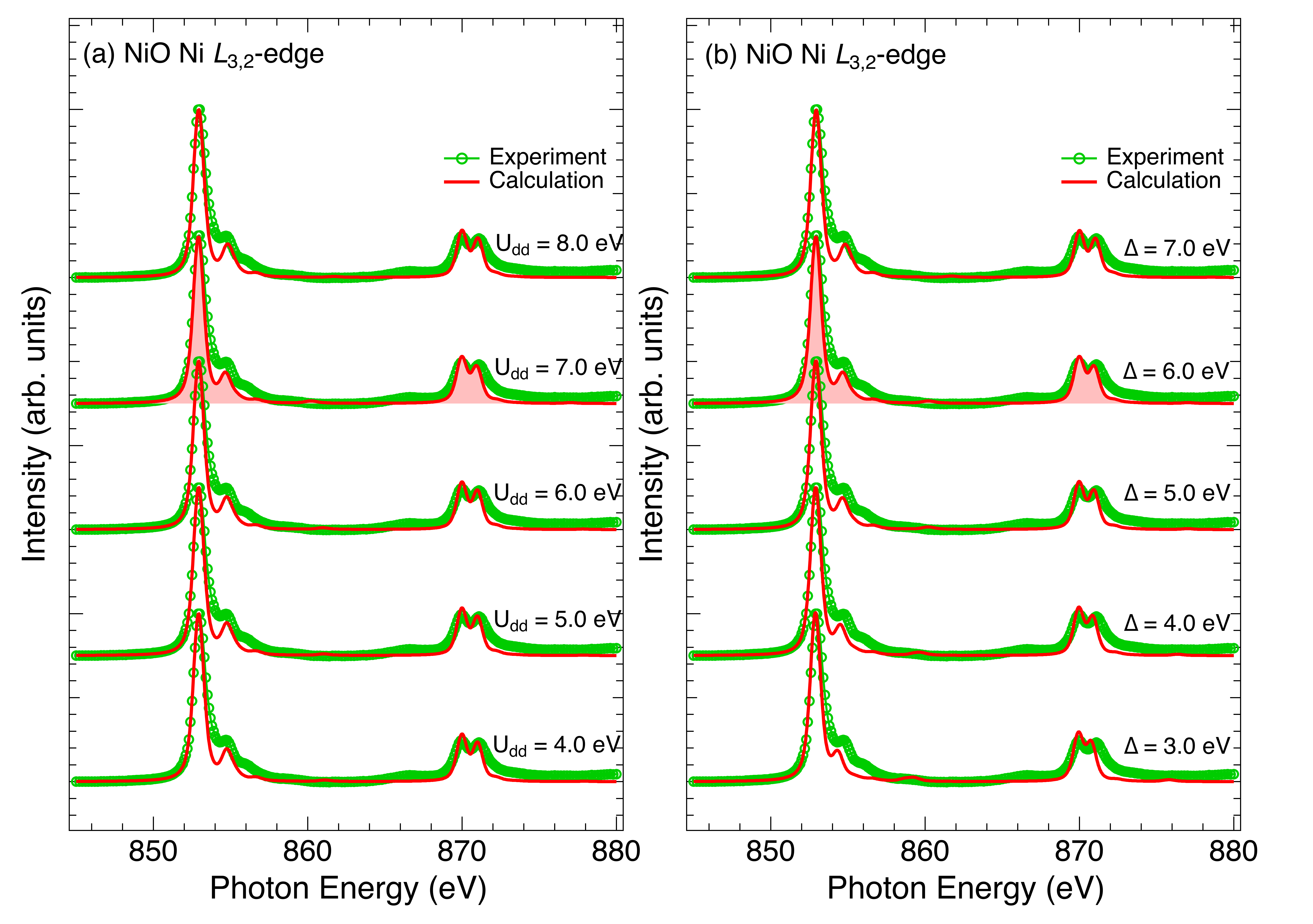}
\caption{(a) Optimization of on-site Coulomb energy $U_{dd}$ and (b) charge transfer energy $\Delta$ for Ni $L_{3,2}$-edge XAS of NiO}
\end{figure}

In Fig. 7(a) and (b), we plot a series of calculated Ni 2$p$ PES and Ni $L$-edge XAS spectra, respectively, in order to check the optimal value of F\textsubscript{k} and G\textsubscript{k} for NiTe$_2$. By varying them together from a reduction factor of R = 1.0 to 0.1, keeping all other parameters fixed to the optimal values for NiTe$_2$, we confirm that for R = 1.0 i.e. no reduction of F\textsubscript{k} and G\textsubscript{k} results in the least deviation compared to experiment.

In Fig. 8(a) and (b), we plot a series of calculated Ni 2$p$ PES and Ni $L$-edge XAS spectra, respectively, to check the optimal value of charge transfer energy $\Delta$ for NiTe$_2$. By varying it from $\Delta$ = -3.5 eV to +2.0 eV, and keeping all other parameters fixed to the optimal values, it is confirmed that the Ni 2$p_{3/2}$ and Ni 2$p_{1/2}$ PES satellites to the main peaks as well as the Ni $L$-edge XAS satellites show the least deviation compared to the experiment for $\Delta$ = -2.8 eV.

Figure 9 (a) and (b) shows a series of calculated Ni $L_{3,2}$-edge XAS spectra for checking the the optimal value of on-site Coulomb energy $U_{dd}$ and charge transfer energy $\Delta$ for NiO, respectively. $U_{dd}$ is varied from 4.0 eV to 8.0 eV, in 1.0 eV steps, by keepng all other parameter fixed (Fig. 10 (a)) and $\Delta$ is varied from 3.0 eV to 7.0 eV, in 1.0 eV steps, by keeping all other parameters fixed (Fig. 10 (b)). The results confirm that the least deviation between calculated and experimental spectra is obtained for $U_{dd}$ = 7.0 eV and $\Delta$ = 6.0 eV.

In order to understand the evolution from an effective positive-$\Delta$ to negative-$\Delta$,  we carried out a set of CT cluster model calculations to investigate the relation of $\Delta$ and $U_{dd}$ with $d^n$. In Fig. 10,  we plot the variation of $d^n$ count vs. $T_{eg}$ for different $\Delta$ values ((red curves), and all other parameters were fixed to optimal values of NiTe$_2$. The curves show that for Ni$^{2+}$, starting with $d^n$ = 8 for $T_{eg}$ = 0, a sharp increase of $\sim$1 electron to $d^n$$\sim$9 is obtained for a finite $T_{eg}$ = 5 meV, for values of negative $\Delta$$\leq$$\Delta_C$ = -1.55 eV. This is attributed to a spontaneous charge transfer which causes an effective negative-$\Delta$ to form the dominantly $3d^{n+1}\underline{L}^1$ ground state even for a small $T_{eg}$ = 5 meV. In region A of Fig. 10, it is seen that using optimal parameters ($\Delta$= -2.8 eV and $T_{eg}$ = 1.8 eV) corresponding to spectra shown in Figs. 6(a,b), we obtain  $d^n$ = 9.1 (black square in red curve for $\Delta$= -2.8 eV). 
Fig. 10 also shows the variation of $d^n$ count vs. $T_{eg}$ for $U_{dd}$ = 3.0 to 6.0 eV (green curves; all other parameters fixed to optimal values of NiTe$_2$). The region A (pink shade) in Fig. 10 shows that an increase in $d^n$ is obtained on reducing $\Delta$ (making it more negative;red arrow) or reducing $U_{dd}$(green arrow). This indicates that a large CT from ligand to Ni site in NiTe$_2$ is obtained from a combination of effective negative-$\Delta$$\leq$$\Delta_C$ and reduced $U_{dd}$. The reduction of $U_{dd}$ is attributed to the larger polarizability
of Te compared to oxygen anions, as polarizability is roughly proportional to the anion size\cite{ZS}. Also, for $\Delta$$>$$\Delta_C$ i.e. $\Delta$ is more positive than 
$\Delta_C$ (region B; purple shade), the sharp jump in $d^n$ for $T_{eg}$ = 5 meV gets suppressed and instead shows a gradual rise as a function of  $T_{eg}$.

\begin{figure}
\includegraphics[scale=0.20]{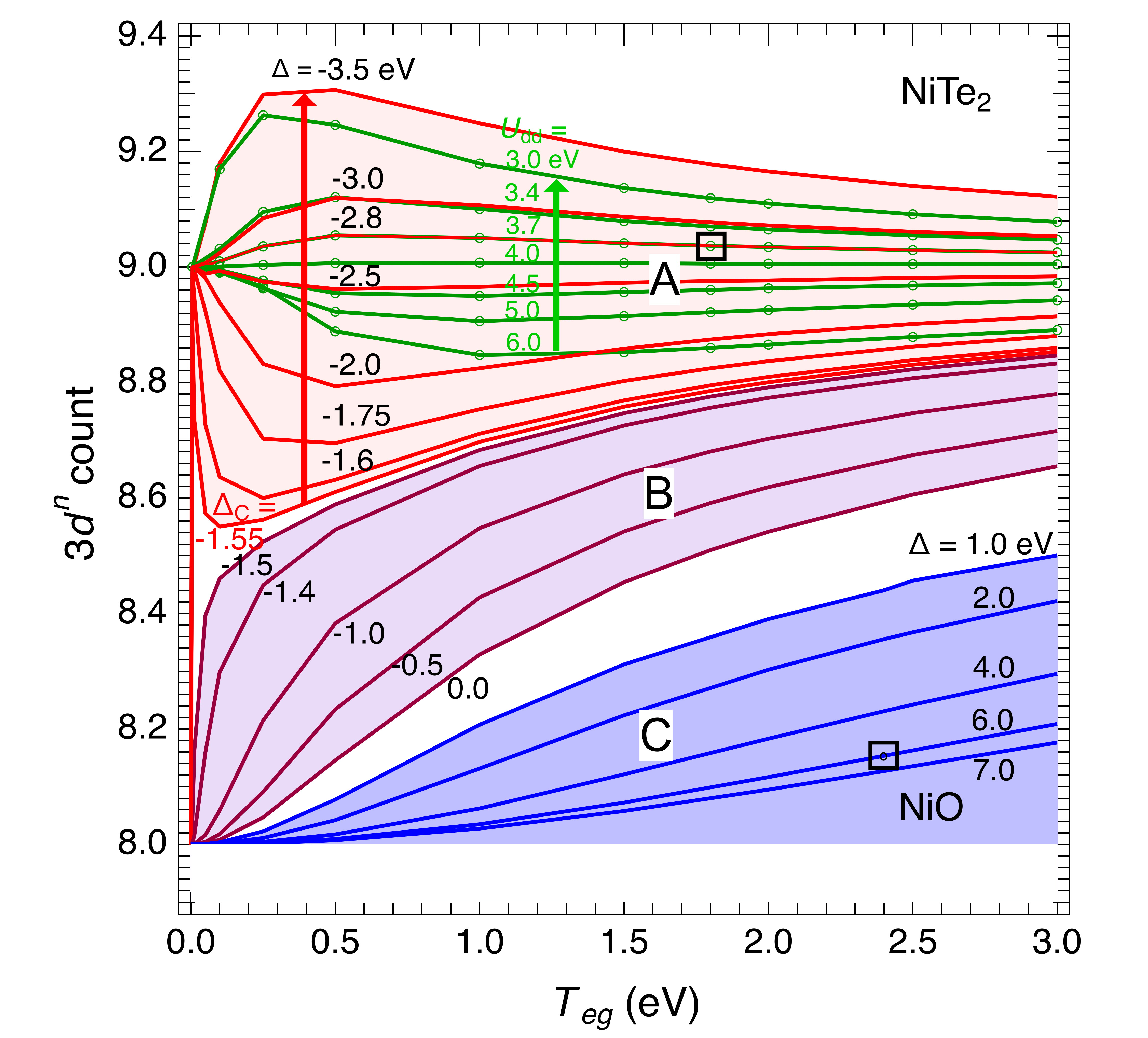}
\caption{Variation of $d^n$-count vs. $T_{eg}$ for selected values of $\Delta$ and $U_{dd}$ of NiTe$_2$ and NiO. Region A corresponds to effective negative-$\Delta$, and regions B and C correspond  of effective positive-$\Delta$. Squares($\square$) indicate optimal values which reproduce experimental spectra shown in Fig. 6.}
\end{figure}

Since it is somewhat surprising  that the negative-$\Delta$ region occurs only for  $\Delta$$<$$\Delta_C$ = -1.55 eV in Fig. 10, we checked the relation of spin magnetic moments with $\Delta$, and also calculated a series of XAS spectra for various $\Delta$ values across $\Delta_C$ = -1.55 eV. 
Figure 11(a), shows the spin magnetic moment $m_S$ vs. $\Delta$ for $T_{eg}$ = 5 meV obtained from the 
same cluster model calculation results shown in Fig. 10, with all other parameters fixed to NiTe$_2$ optimal parameters. The $m_S$ values exhibit a jump at $\Delta$ = -1.55 eV, where even the total $d^n$ count shows a jump in Fig. 10, and we denoted this value as $\Delta_C$. The $m_S$ values show negligible change for $\Delta$$\leq$$\Delta_C$, and also for $\Delta$$>$$\Delta_C$. 
However, the $m_S$ vs. $\Delta$ curve for the optimal value of $T_{eg}$ = 1.2 eV for NiTe$_2$,  $m_S$ shows a small gradual increase on increasing $\Delta$.
The corresponding individual ground state weights from the cluster model calculations for $T_{eg}$ = 5 meV are listed in Table IV.
It is clear from Fig. 10 and Table IV that for $\Delta$$\leq$$\Delta_C$, the results indicate a dominantly
 $d^{n+1}\underline{L}^1$ state corresponds to the effective negative-$\Delta$ region A of Fig. 10, and for 
$\Delta$$>$$\Delta_C$, it indicates a dominantly  $d^{n}$ state and corresponds to the effective positive-$\Delta$ region B of Fig. 10.

\begin{table}[t!]
	
	 \caption{Ground state weights (\%) obtained from cluster model calculations with $T_{eg}$ = 5 meV, starting with the formal $d^8$ configuration for NiTe$_2$}
\begin{tabular}{cccc}

\hline

NiTe$_2$					&$d^8$&$d^{9}\underline{L}^1$&$d^{10}\underline{L}^2$\\
\hline 
$\Delta$ = $\Delta_C$ = -1.55 eV		&~10.3\%~&~89.7\%~&~0.0\%~	\\
(and all $\Delta$$<\Delta_C$) 				&		&		&			\\
\hline
$\Delta$ =	 -1.5 eV			&~95.8\%~&~0.4.2\%~&~0.0\%~			\\
(and all $\Delta$ more +ve than $\Delta_C$)				&		&		&			\\
\hline \\

\end{tabular}

\end{table}

\begin{figure}
\includegraphics[scale=0.20]{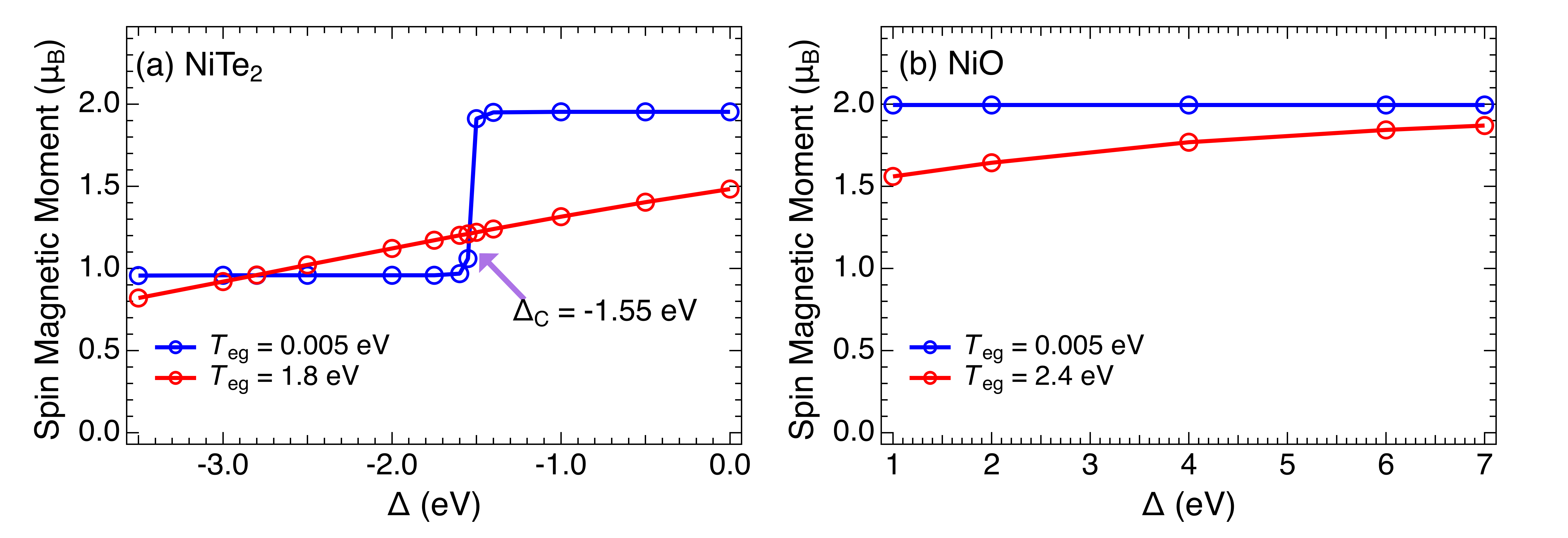}
\caption{(a) In NiTe$_2$, the spin magnetic moment $m_S$ vs. $\Delta$ for $T_{eg}$ = 5 meV, exhibits a jump at $\Delta_C$ = -1.55 eV, while for optimal $T_{eg}$ = 1.8 eV,  $m_S$ vs. $\Delta$ shows a gradual change across $\Delta_C$.
 (b) In NiO, the $m_s$ vs. $\Delta$ shows a nearly constant value for small $T_{eg}$ = 5 meV, while for optimal $T_{eg}$ = 2.5 eV,  $m_S$ vs. $\Delta$ shows a gradual change with  
 $\Delta$.}
\end{figure}

Similarly, for NiO, the spin magnetic moment $m_S$ vs. $\Delta$ for $T_{eg}$ = 5 meV are plotted in Fig. 11(b). These are also obtained from the calculation results of NiO shown in Fig. 10 (region C; blue curves; blue shade). The NiO results show that the $d^n$ count starts from  $d^n$ = 8 for $T_{eg}$ = 0, and shows a small gradual increase on increasing $T_{eg}$. Further, as a function of $\Delta$, the the $d^n$ count shows a systematic decrease on increasing $\Delta$ for fixed $T_{eg}$ values. The $m_S$ values vs. $\Delta$ for $T_{eg}$ = 5 meV show a negligible change for all $\Delta$ values, while the $m_S$ vs. $\Delta$ curve for the optimal value of $T_{eg}$ = 2.4 eV shows a small gradual increase of $m_S$ on increasing $\Delta$. These results indicate that region C of Fig. 10 corresponds to the positive-$\Delta$ region, with qualitatively similar curves as in region B. 
In particular, for optimal parameters of NiO ($\Delta$= 6.0 eV and $T_{eg}$ = 2.4 eV) corresponding to spectra shown in Fig. 6(c), the obtained value of  $d^n$ = 9.1
(black square in blue curve for $\Delta$= 6.0 eV in region C of Fig. 10).
Further, the $m_S$ values on increasing $\Delta$ in Fig. 11(a) connect to the $m_S$ values on increasing $\Delta$ in Fig. 11(b), thus confirming that region B is an effective positive-$\Delta$ region.

The above observations are qualitatively similar to the behavior of spin magnetic moments $m_S$ vs. $\Delta$ in CoTe$_2$. In order to answer the question about why $\Delta_C$ = -1.55 eV in NiTe$_2$, we follow the case of CoTe$_2$, for which we have recently reported\cite{ShelkeCoTe2} that regions A and B are separated across $\Delta_C$ = 0.3 eV. It was found that the true negative-$\Delta$ region is obtained when the charge transfer energy is defined as the energy difference between the energies of the lowest multiplet of $d^{n+1}\underline{L}^1$ and $d^n$  configurations\cite{Bocquet,Fujimori,Fujimori2}, and not in terms of the difference between the
average energy of the $d^n$ multiplets and the average energy of $d^{n+1}\underline{L}^1$ multiplets.
  
Accordingly, we then checked at what value does the ground state transform from a dominantly $d^{n}$ state to $d^{n+1}\underline{L}^1$ state i.e. negative $\Delta$ represents $\Delta$$<0$ or $\Delta_{eff}$$<0$ for NiTe$_2$.
 In cluster model calculations,  $\Delta$ is usually defined as,
$\Delta$ = $E$($d^{n+1}\underline{L}^1$)- $E$($d^n$), 
where $E$($d^n$) is the average energy of the $d^n$ multiplets and $E$($d^{n+1}\underline{L}^1$) is the average energy of $d^{n+1}\underline{L}^1$ multiplets. 
On the other hand, if 
the energy difference of the lowest multiplet and the average multiplet energy is $\Delta'$$E_n$, then the effective charge transfer energy is
 $\Delta_{eff}$ = $\Delta$ + $\Delta'$$E_{n+1}$ - $\Delta'$$E_n$.
We calculated $L$-edge XAS spectra with a small $T_{eg}$ = 5 meV for various $\Delta$ values at and across $\Delta_C$, with a very small Gaussian broadening  of 0.1 eV FWHM as shown in Fig. 12.
The $L$-edge XAS spectrum corresponds to transitions from the ground state to final states of the type
$2p^63d^n$$\rightarrow$$2p^53d^{n+1}$, $2p^63d^{n+1}\underline{L}^1$$\rightarrow$$2p^53d^{n+2}\underline{L}^1$, etc. 
The obtained results shown in Fig. 12 show that
the actual transition to a negative charge-transfer character takes place when $\Delta_{eff}$$<0$, and not for $\Delta$$<$0, as described in the following.

\begin{figure}
\includegraphics[scale=0.70]{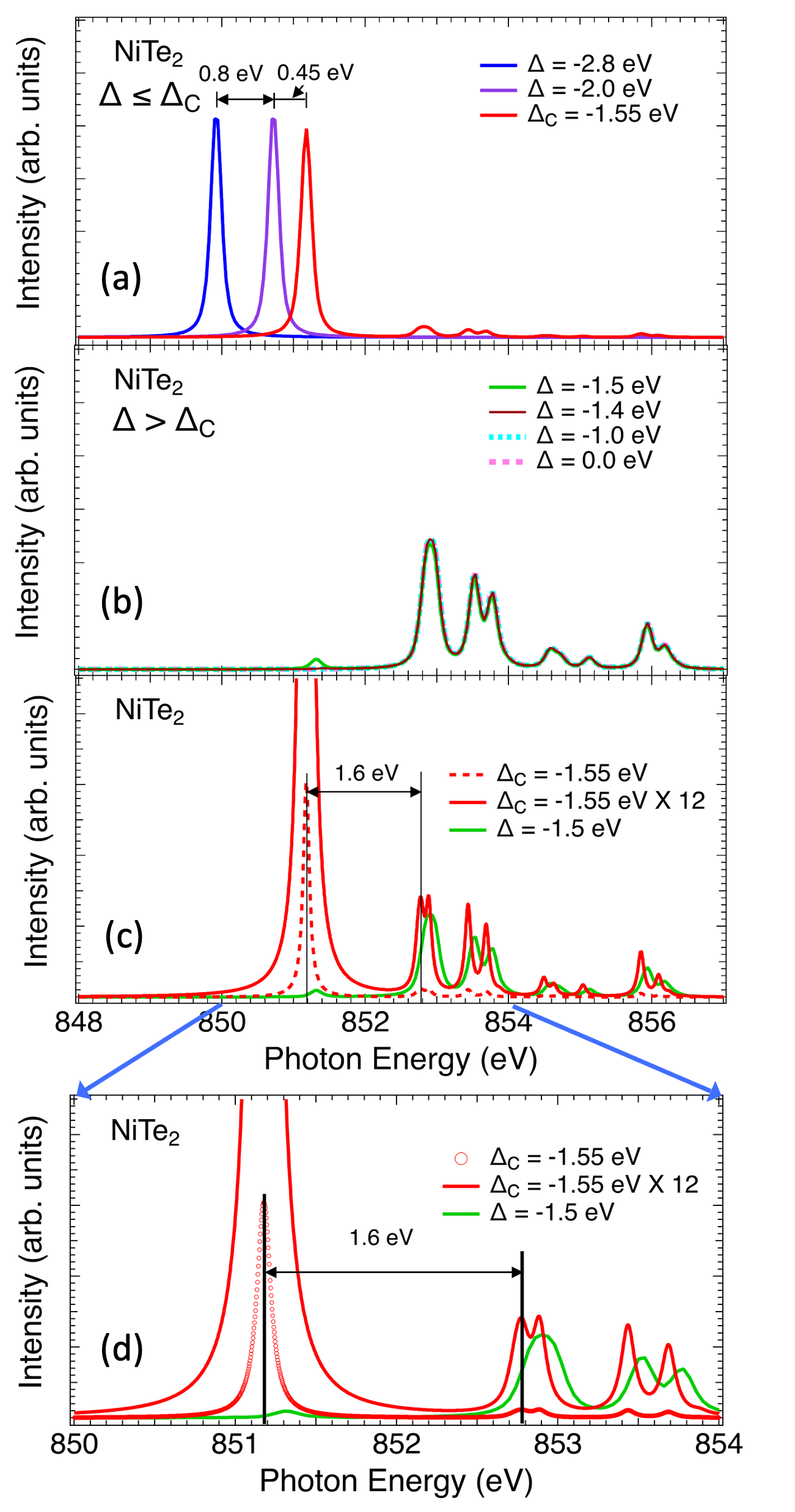}
\caption{ The XAS spectral calculations for $T_{eg}$ = 5 meV and varying $\Delta$ with all other parameters fixed to optimal values of NiTe$_2$: (a) for $\Delta$$\leq$$\Delta_C$ = -1.55 eV, (b) for $\Delta$$>$$\Delta_C$ = -1.55 eV (c) for $\Delta$ = $\Delta_C$ = -1.55 eV ($d^{n+1}\underline{L}^1$ state) and $\Delta$ = -1.5 eV ($d^{n}$ state) (d) same as (c), but plotted on an expanded x-scale.}
\end{figure}

Fig. 12(a-c) shows the $L$-edge XAS spectra with a small $T_{eg}$ = 5 meV and a small Gaussian broadening  of 0.1 eV FWHM for selected $\Delta$ values, with all other parameters fixed to optimal values of NiTe$_2$
The spectra in panel(a) for $\Delta$$\leq$$\Delta_C$ = -1.55 eV  show hardly any change in the shape of the multiplet features but show a systematic shift equal to the change in $\Delta$. From Table IV, it is clear that all the spectra  originate in the dominantly $d^{n+1}\underline{L}^1$ initial state multiplets.
In panel (b), we plot four spectra for $\Delta$$>$$\Delta_C$ = -1.55 eV (i.e. $\Delta$ more positive than $\Delta_C$) and the results show very similar spectra
but with significantly different multiplet features.
From Table IV, we know that for $\Delta$ = 1.5 eV, the initial state is dominated by the $d^{n}$ state. This indicates the spectra in panel (b) originate in the $d^{n}$ state and do not show a shift as is seen in Fig. 12(a) for $\Delta$$\leq$$\Delta_C$. 
In panel(c), we compare the spectra for $\Delta_C$ = -1.55 eV with $\Delta$ = -1.5 eV. We also plot the spectrum of $\Delta$ = -1.55 eV calculated with a smaller 0.01 eV FWHM Gaussian broadening to see fine features. Panel (d) shows the same spectra plotted on an expanded x-scale. The lowest energy multiplet of the $\Delta_C$ = -1.55 eV spectrum is shifted by 1.6$\pm$0.1 eV compared to the dominant energy multiplet of the $\Delta$ = -1.5 eV spectrum. 
This difference of the spectral behavior for $\Delta$$\leq$$\Delta_C$ and $\Delta$$>$$\Delta_C$, together with the results of $d^n$ count of Fig. 10 indicate that region A with $\Delta$$\leq$$\Delta_C$ = -1.55 eV corresponds to the effective negative-$\Delta$ region, and region B with $\Delta$$>$$\Delta_C$ corresponds to an effective positive-$\Delta$ region. 

We have thus confirmed that $\Delta_C$ for NiTe$_2$ corresponds to attaining an effective negative $\Delta$, defined as the energy difference between lowest multiplet of $d^n$ and $d^{n+1}\underline{L}^1$  states\cite{Bocquet,Fujimori,Fujimori2}, i.e. a material attains a genuine negative-$\Delta$ state for $\Delta$$\leq$$\Delta_C$, when the lowest multiplet of the $d^{n+1}\underline{L}^1$ state becomes more negative than lowest multiplet of the $d^{n}$ state. Fig. 10 highlights regions of effective negative-$\Delta$ (A) and effective positive-$\Delta$ (B, C). A similar detailed analysis of the relation of $d^n$ with $\Delta$ and $U_{dd}$ in CoTe$_2$ using CT cluster model calculations is reported in ref.\cite{ShelkeCoTe2}.

\begin{figure}
\includegraphics[scale=0.08]{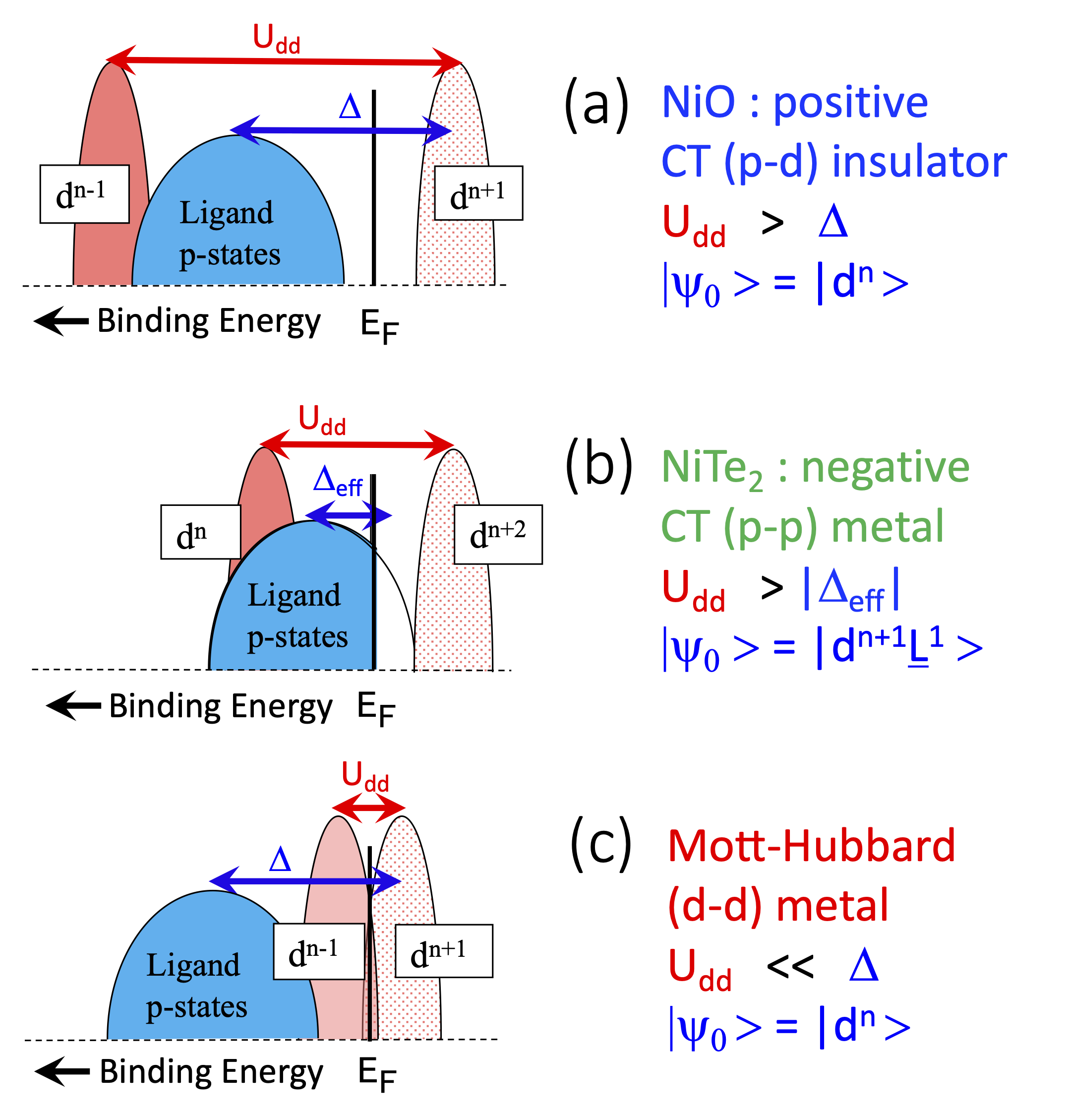}
\caption{Schematic electronic structure of: 
(a) NiO: a positive-$\Delta$ charge-transfer insulator with $U_{dd}$$>$$\Delta$ and 
\textit{p}$\rightarrow$\textit{d} type lowest energy excitations
(b) NiTe$_2$: an effective negative-$\Delta$ metal with a reduced $U_{dd}$, but with $U_{dd}$$>$$|\Delta_{eff}|$ and $E_F$ is within the ligand-$p$ band states, facilitating band inversion with \textit{p}$\rightarrow$\textit{p} type lowest energy excitations to make it a correlated topological metal.
(c) a Mott-Hubbard metal with $U_{dd}$$<<$$\Delta$. A slightly larger $U_{dd}$ but still less than $\Delta$ would result in a Mott-Hubbard insulator if the lower(occupied) and upper(unoccupied) Hubbard $d$-bands have a band gap.
For Mott-Hubbard and positive-$\Delta$ materials, the ground state $|\psi_0$$>$ has dominantly $|d^{n}$$>$ character, while for negative-$\Delta$ materials, $|\psi_0$$>$ has dominantly $|d^{n+1}\underline{L}^1$$>$ character, where $\underline{L}^1$ is a ligand $p$-band hole.
}
\end{figure}

 Figure 13(a) shows the schematic electronic structure of NiO with $U_{dd}$$>$$\Delta$. It is noted that while the NiO Ni $L$-edge XAS can be reproduced fairly with a single metal site cluster model calculation\cite{Veenendaal, alders1998}, the Ni 2p core level PES was shown to require a multiple Ni-site calculation\cite{VeenendaalPRL}, or a single metal-site CT multiplet calculation combined with DMFT calculations\cite{Ghiasi}, to reproduce the additional well-screened peak observed in 2p PES arising from non-local screening. Fig. 13(b) shows the schematic electronic structure of NiTe$_2$ with a reduced $U_{dd}$ compared to NiO, but still $U_{dd}$$>$$|\Delta|$ with a dominantly $d^{n+1}\underline{L}^1$ contribution in the ground state.

The obtained values of $d$-$p$ hybridization strength $T_{eg}$ indicate that $T_{eg}$ is smaller for NiTe$_2$ compared to NiO, consistent with Ni-Te distance (= 2.620$\AA$)  being larger than Ni-O distance (= 2.09$\AA$).
 This indicates that the reduced $U_{dd}$ in NiTe\textsubscript{2} compared to NiO is not due to an increase in $T_{eg}$. The large increase in $d^n$ count on the Ni site ($d^n$ = 9.1) by more than one electron compared to a formal $d^n$ = 8 for divalent NiTe$_{2}$ is attributed to negative-$\Delta$ and a reduced $U_{dd}$.  While the value of $U_{dd}$ in NiTe$_2$ gets reduced by nearly 50\% compared to NiO,
 the moderately repulsive value of $U_{dd}$ = 3.7 eV is crucial to achieve topological properties in NiTe$_2$ as follows: if $U_{dd}$ was $<<$$|\Delta|$, NiTe$_2$ would exhibit a Mott-Hubbard metal character with \textit{d}$\rightarrow$\textit{d} type lowest energy excitations (Fig. 13(c)) in the ZSA phase diagram\cite{zaanen1985}. This would imply weak spin-orbit coupled 3$d$ states in the vicinity of $E_F$ and absence of band inversion. Accordingly, NiTe$_2$ would not be a topological Dirac semi-metal.
Thus, only because $U_{dd}$$>$$|\Delta|$, the 3$d$ states are pushed away from $E_F$ and NiTe$_{2}$ becomes a topological Dirac semi-metal with band inversion and \textit{p}$\rightarrow$\textit{p} type excitations between strongly spin-orbit coupled Te $5p$ derived states.
The results then show that the present approach, based on quantifying the value of $U_{dd}$ using the Cini-Sawatzky method from experimental single particle DOS and the two hole satellite energy provides a valid estimate of $U_{dd}$. Having employed this value in cluster model calculations to obtain theoretically simulated spectra which are close to the experimental spectra, we obtain the additional parameters of a negative charge transfer energy $\Delta$ and $T_{eg}$ for NiTe$_2$. From this exercise, our results show that a finite $U_{dd}$ is indeed necessary to explain the electronic structure of NiTe$_2$ (moderately correlated negative-$\Delta$), as well as its comparison with NiO (strongly correlated positive-$\Delta$)  in the ZSA phase diagram, and provides a more complete picture compared to earlier work.

\section{\label{sec:levelII}Conclusions}
In conclusion, the core-level and valence band electronic structure of single crystal NiTe$_2$ was investigated for quantifying electronic parameters in NiTe$_2$. Using the Cini-Sawatzky method, we obtain a value of $U_{dd}$ = 3.7 eV. The Ni $2p$ core level and $L$-edge XAS spectra were analyzed by charge transfer cluster model calculations using the obtained $U_{dd}$ (= 3.7 eV), and the results indicate NiTe$_2$ is a negative charge-transfer  material with $\Delta$ = -2.8 eV. The same type of cluster model analysis for NiO $L$-edge XAS confirms its well-known strongly correlated charge-transfer insulator character, with $U_{dd}$ = 7.0 eV and $\Delta$ = 6.0 eV. The hybridization strength $T_{eg}$ between Ni $3d$ and ligand states for NiTe$_2$$<$NiO, and indicates that the reduced $U_{dd}$ in NiTe\textsubscript{2} compared to NiO is not due to an increase in $T_{eg}$. The $d^n$ count on the Ni site increases by nearly one electron in NiTe$_{2}$ due to negative-$\Delta$ and a reduced $U_{dd}$. Since $U_{dd}$$>$$|\Delta|$, the results indicate the important requirement of a finite repulsive $U_{dd}$ in making NiTe$_{2}$ a moderately correlated $p$-type Dirac semi-metal.

\begin{acknowledgments}
This work was supported by the National Science and Technology Council(NSTC) of Taiwan under Grant Nos. NSTC 113-2112-M-006-009-MY2 (CNK), 110-2124-M-006-006-MY3 (CSL), 112-2124-M-006-009 (CSL), 113-2112-M-007-033 (AF),  112-2112-M-213-029(AC) and 114-2112-M-213-021(AC). AF acknowledges support from the Yushan Fellow Program under the Ministry of Education of Taiwan and Grant No. JP22K03535 from Japan Society for the Promotion of Science(JSPS).  ARS thanks the National Science and Technology Council(NSTC) of Taiwan for a post-doctoral fellowship under Grant No. NSTC 114-2811-M-213-006.  
\end{acknowledgments}

\end{document}